\documentclass[10pt,aps,prc,floatfix,twocolumn,nofootinbib,superscriptaddress]{revtex4-2}
\usepackage{graphicx,amsmath,amssymb,bm}
\usepackage{amsfonts}
\usepackage[utf8]{inputenc}
\usepackage{verbatim}
\usepackage{float}
\usepackage[labelfont={small},font=small,subrefformat=parens,caption=false]{subfig}
\usepackage{cancel}
\usepackage{multirow}
\usepackage{array}
\usepackage{xparse}
\usepackage{xspace}
\usepackage{esint}

\usepackage{bigstrut}
\usepackage{multirow, boldline}
\usepackage{adjustbox}
\usepackage{comment}
\usepackage{amssymb}

\usepackage{braket}
\newcommand{\dv}[2]{\frac{\textup{d}#1}{\textup{d}#2}}
\usepackage{color}
\usepackage{dsfont}
\usepackage{dcolumn}

\usepackage[pdfencoding=auto, pdfpagelabels]{hyperref}

\usepackage[overload]{textcase}

\definecolor{linkcolor}{rgb}{0,0,0.40} 
\hypersetup{%
    pdfsubject=Paper,
    pdfkeywords={nuclear physics} {Bayesian} {chiral EFT} {emulator} {variational principle},
    unicode = true,
    breaklinks = true,
    colorlinks = true,
    linkcolor = linkcolor,
    citecolor = linkcolor,
    menucolor = linkcolor,
    urlcolor = linkcolor
}

\usepackage{booktabs}
\usepackage{cellspace}
\graphicspath{{./figures/}}

\makeatletter
\newcommand\newsubcommand[3]{\newcommand#1{#2\sc@sub{#3}}}
\def\sc@sub#1{\def\sc@thesub{#1}\@ifnextchar_{\sc@mergesubs}{_{\sc@thesub}}}
\def\sc@mergesubs_#1{_{\sc@thesub#1}}

\newcommand\newsupcommand[3]{\newcommand#1{#2\sc@sup{#3}}}
\def\sc@sup#1{\def\sc@thesup{#1}\@ifnextchar^{\sc@mergesups}{^{\sc@thesup}}}
\def\sc@mergesups^#1{^{\sc@thesup#1}}
\makeatother

\DeclareMathAlphabet{\mathbcal}{OMS}{cmsy}{b}{n}

\newcommand{\etal}{\textit{et~al.}\xspace}

\newcommand{\abinitio}{\textit{ab~initio}\xspace}

\newcommand{\fmi}{\, \text{fm}^{-1}}

\newcommand{\MeV}{\, \text{MeV}}

\newcommand{\NNNLO}{\ensuremath{{\rm N}{}^3{\rm LO}}\xspace}

\newcommand{\param}{\boldsymbol{\theta}}
\newcommand{\nlecs}{n_a}
\newcommand{\nbasis}{n_b}
\newcommand{\weights}{\beta}
\newcommand{\kvpweights}{\vec{\weights}_\star}
\newcommand{\lagmult}{\lambda_\star}
\newcommand{\psitrial}{\widetilde \psi}
\newcommand{\dU}{\Delta \widetilde U}
\newcommand{\genkvp}{\mathcal{L}}

\newcommand{\npr}{\ensuremath{np}}

\newcommand{\ordervec}{\vec}

\newcommand{\inputvec}{\mathbf}

\newsubcommand{\ckvec}{\ordervec{c}}{k}

\newsubcommand{\bkvec}{\ordervec{b}}{k}
\newsubcommand{\ckvecset}{\ordervec{\inputvec{c}}}{k}

\newsubcommand{\ckvecapprox}{\mathbf{c}'}{k}
\newsubcommand{\ckvecapproxset}{\mathbf{C}'}{k}

\newsubcommand{\bkvecapprox}{\mathbf{b}'}{k}
\newsubcommand{\bkvecset}{\mathbf{B}}{k}
\newsubcommand{\bkvecapproxset}{\mathbf{B}'}{k}

\newcommand{\genobs}{y}

\newsubcommand{\genobsvec}{\ordervec{\genobs}}{k}
\newsubcommand{\genobsvecset}{\ordervec{\inputvec{\genobs}}}{k}

\newsubcommand{\akvec}{\mathbf{a}}{k}

\newsubcommand{\akvecapprox}{\mathbf{a}'}{k}
\newsubcommand{\akvecset}{\mathbf{A}}{k}
\newsubcommand{\akvecapproxset}{\mathbf{A}'}{k}

{}  % Remove the definition from the Physics package

\newcommand{\Elab}{E_{\rm lab}}

\newcommand{\kzero}{k_0}

\newcommand{\trans}{\intercal}

\newcommand{\chiEFT}{$\chi$EFT}

\def\diffd{\mathrm{d}}  % Upright differentials
\DeclareDocumentCommand\differential{ o g d() }{ % Differential 'd'
    \IfNoValueTF{#2}{
        \IfNoValueTF{#3}
            {\diffd\IfNoValueTF{#1}{}{^{#1}}}
            {\mathinner{\diffd\IfNoValueTF{#1}{}{^{#1}}\argopen(#3\argclose)}}
        }
        {\mathinner{\diffd\IfNoValueTF{#1}{}{^{#1}}#2} \IfNoValueTF{#3}{}{(#3)}}
    }
\DeclareDocumentCommand\dd{}{\differential} % Shorthand for \differential

\newcommand{\pathd}{\mathcal{D}}  % differential symbol for path integrals

\DeclareDocumentCommand\pathdifferential{ o g d() }{ % Path 'D'
    \IfNoValueTF{#2}{
        \IfNoValueTF{#3}
            {\pathd\IfNoValueTF{#1}{}{^{#1}}}
            {\mathinner{\pathd\IfNoValueTF{#1}{}{^{#1}}\argopen(#3\argclose)}}
        }
        {\mathinner{\pathd\IfNoValueTF{#1}{}{^{#1}}#2} \IfNoValueTF{#3}{}{(#3)}}
    }

\newcommand{\orcid}[1]{\href{https://orcid.org/#1}{\includegraphics[scale=0.055]{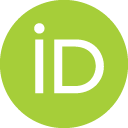}}}

\begin{document}

\title{Wave-function-based emulation for nucleon-nucleon scattering in momentum space}

\author{A.~J. Garcia \orcid{0000-0003-1723-3225}}
\email{garcia.823@osu.edu}
\affiliation{Department of Physics, The Ohio State University, Columbus, Ohio
43210, USA}

\author{C. Drischler \orcid{0000-0003-1534-6285}}
\email{drischler@ohio.edu}
\affiliation{Department of Physics and Astronomy and Institute of Nuclear and Particle Physics, Ohio University, Athens, Ohio 45701, USA}
\affiliation{Facility for Rare Isotope Beams, Michigan State University, MI 48824, USA}

\author{R.~J. Furnstahl \orcid{0000-0002-3483-333X}}
\email{furnstahl.1@osu.edu}
\affiliation{Department of Physics, The Ohio State University, Columbus, Ohio 43210, USA}

\author{J.~A. Melendez \orcid{0000-0003-1359-1594}}
\email{melendez.27@osu.edu}
\affiliation{Department of Physics, The Ohio State University, Columbus, Ohio 43210, USA}

\author{Xilin Zhang \orcid{0000-0001-9278-5359}}
\email{zhangx@frib.msu.edu}
\affiliation{Facility for Rare Isotope Beams, Michigan State University, Michigan 48824, USA}

\newcommand{\verifyvalue}[1]{#1}

\newcommand{\umatrix}{\boldsymbol{u}}
\newcommand{\indexfunct}{\alpha}

% rjf
\newcommand{\psitest}[4]{#1\psi_{#4}^{#2}#3}
\newcommand{\mypsitest}{\psitest{(}{\alpha}{)}{t}}  % use \mypsitest{}^{ss'}
\newcommand{\mypsitestpp}{\psitest{}{}{}{}}   %  \mypsitestpp{}^{ss'}

% ajg
\newcommand{\psitestalt}[3]{#1\psi^{#2}#3}
\newcommand{\mypsitestalt}{\psitestalt{(\widetilde}{\alpha}{)}}  % use \mypsitestalt{}^{ss'}

\date{\today}

%%%%%%%%%%%%%%%%%%%%%%%%%%%%%%%%%%%%%%%%%%%%%%%%%%%%%%%%
\begin{abstract}
Emulators for low-energy nuclear physics can provide fast \& accurate predictions of bound-state and scattering observables for applications that require repeated calculations with different parameters, such as Bayesian uncertainty quantification. In this paper, we extend a scattering emulator based on the Kohn variational principle (KVP) to momentum space (including coupled channels) with arbitrary boundary conditions, which enable the mitigation of spurious singularities known as Kohn anomalies. We test it on a modern chiral nucleon-nucleon ($NN$) interaction, including emulation of the coupled channels. We provide comparisons between a Lippmann-Schwinger equation emulator and our KVP momentum-space emulator for a representative set of neutron-proton (\npr) scattering observables, and also introduce a quasi-spline-based approach for the KVP-based emulator. Our findings show that while there are some trade-offs between accuracy and speed, all three emulators perform well. Self-contained Jupyter notebooks that generate the results and figures in this paper are publicly available.

\end{abstract}
%%%%%%%%%%%%%%%%%%%%%%%%%%%%%%%%%%%%%%%%%%%%%%%%%%%%%%%%

\maketitle
%%%%%%%%%%%%%%%%%%%%%%%%%%%%%%%%%%%%%%%%%%%%%%%%%%%%%%%%

%%%%%%%%%%%%%%%%%%%%%%%%%%%%%%%%%%%%%%%%%%%%%%%%%%%%%%%%
\section{Introduction} \label{sec:intro}
%%%%%%%%%%%%%%%%%%%%%%%%%%%%%%%%%%%%%%%%%%%%%%%%%%%%%%%%

Nucleon-nucleon ($NN$) scattering has long been used to fix parameters of microscopic Hamiltonians designed for \abinitio few- and many-body calculations.
But the uncertainty in most existing nuclear models has been underestimated because they have lacked two key ingredients: a rigorous accounting of Hamiltonian uncertainty and a complete estimate of parameter uncertainty.

In the case of chiral effective field theory (\chiEFT)~\cite{Epelbaum:2008ga, Machleidt:2011zz, Hammer:2019poc, Epelbaum:2019kcf}, Hamiltonian uncertainty manifests as a truncation error, which has been statistically modeled in Refs.~\cite{Furnstahl:2015rha,Melendez:2017phj,Melendez:2019izc,Melendez:2020xcs}.
A holistic parameter estimation study would then both account for truncation errors in the likelihood, and estimate and propagate all plausible values of the low-energy constants (LECs) rather than finding a single parameter value maximizing the likelihood.
Bayesian statistical methods are particularly suitable for these tasks~\cite{Higdon:2014tva, Wesolowski:2018lzj,Phillips:2020dmw,Melendez:2020ikd, Svensson:2022kkj}, but are computationally demanding, especially when generalizing to include few-body forces.
Emulators---surrogate models that allow for fast \& accurate (but approximate) model predictions---have the potential to alleviate some of these demands~\cite{Drischler:2022yfb}.
In this paper, we extend our recent explorations of emulators for $NN$ scattering~\cite{Furnstahl:2020abp, Drischler:2021qoy, Melendez:2021lyq} to momentum-space wave functions and coupled channels, and test against a representative set 
of neutron-proton (\npr) scattering observables.

The demand for emulators has led nuclear physics to the general field of parametric model order reduction (PMOR), where the goal is to extract the relevant information from a model while reducing the computational cost significantly. An efficient offline-online decomposition is crucial to construct an efficient emulator.
In the offline stage, the emulator is trained with high-fidelity calculations\footnote{Following the terminology of Ref.~\cite{BastosDiagnosticsGaussianProcess2009}, we will refer to the calculational machinery that generates high-fidelity solutions (e.g., LS equation solver) as a simulator.} for selected sets of parameters, also known as snapshots, while making predictions for any other set of parameters are performed in the online stage. The end result is a reduced-order model (ROM) that serves as an emulator. For general overviews of the literature on PMOR techniques and their applications, we refer the reader to Refs.~\cite{Melendez:2022kid, Bonilla:2022rph}. A pedagogical introduction to projection-based emulators for both scattering and bound-state calculations, including interactive, open-source \texttt{PYTHON} code, can be found in Ref.~\cite{Drischler:2022ipa}.

A particular snapshot-based ROM known as the reduced basis method (RBM)\footnote{The RBM has been rediscovered in the low-energy nuclear theory community as eigenvector continuation (EC). See Ref.~\cite{Melendez:2022kid} for more details.} has emerged as an efficient emulator for the prediction of both bound state and scattering observables~\cite{Frame:2017fah, Konig:2019adq, Furnstahl:2020abp}. The foundation of the first emulators for scattering is the Kohn variational principle (KVP) (e.g., for the $K$~matrix), whose snapshots are based on scattering solutions to the Schr{\"o}dinger equation~\cite{Kohn:1948col,taylor2006scattering}. It has been demonstrated for a variety of real and optical potentials that such emulators can be trained for two- and three-body\footnote{In Ref.~\cite{Zhang:2021jmi}, the offline training stage involves calculations in both momentum and coordinate space.} scattering in coordinate space, then evaluated in the form of matrix inversions with low-dimensional matrices~\cite{Furnstahl:2020abp, Drischler:2021qoy, Zhang:2021jmi}.

Subsequently, an emulator of the Lippmann-Schwinger (LS) equation using the Newton variational principle (NVP)~\cite{newton2002scattering} was introduced in Ref.~\cite{Melendez:2021lyq}. In contrast to the KVP emulator, the variational trial basis is composed of scattering matrices (e.g., $K$~matrices) rather than scattering wave functions. Both approaches were shown to quickly and accurately predict the \npr\ phase shifts from a chiral Hamiltonian across a range of parameter values.
In this paper, we compare a momentum-space KVP-based emulator, including emulation of coupled channels and allowing for arbitrary boundary conditions, to the NVP emulator for a representative set of \npr\ observables. For a comparison of the KVP and NVP emulators in a Galerkin framework and a survey on other emulators see Ref.~\cite{Drischler:2022ipa}.

The paper is organized as follows. In Sec.~\ref{sec:formalism}, we review the underlying formalism of the KVP emulators and its extension to momentum space and coupled channels. We then show results for the momentum-space KVP emulator and compare them to the $K$~matrix (NVP) emulator in Sec.~\ref{sec:results}. We demonstrate that spurious singularities known as Kohn (or Schwartz) anomalies~\cite{PhysRev.124.1468,nesbet1980variational} are mitigated using methods from Ref.~\cite{Drischler:2021qoy}. Section~\ref{sec:summary} has a summary and outlook and additional details of the implementation are given in several appendices. The self-contained set of codes that generate all results and figures shown in this paper is publicly available~\cite{BUQEYEsoftware}.

%%%%%%%%%%%%%%%%%%%%%%%%%%%%%%%%%%%%%%%%%%%%%%%%%%%%%%%%
\section{Formalism} \label{sec:formalism}
%%%%%%%%%%%%%%%%%%%%%%%%%%%%%%%%%%%%%%%%%%%%%%%%%%%%%%%%

%%%%%%%%%%%%%%%%%%%%%%%%%%%%%%%%%%%%%%%%%%%%%%%%%%%%%%%%
\begin{table}[tb]
\renewcommand{\arraystretch}{1.2}
\caption{
Notation used in this work.
}
\label{tab:notation}
\begin{ruledtabular}
\begin{tabular}{lp{6.4cm}}
Notation & Description \\ 
% \colrule
\midrule
$\param$ & vector of parameters; $\param_i$ are the parameters for the $i\mathrm{th}$ snapshot \\
$s, s'$ & indices for the exit and entrance channels of the scattering process, e.g., ${^3}S_1$ and ${^3}D_1$\\
$t, t'$ & indices for available channels (summation convention implied) \\
$\psi_i^s$ & wave function in the channel $s$ used for training and associated with the $i\mathrm{th}$ snapshot with $\param_i$ [high-fidelity solution of Eq.~\eqref{eq:schrodinger}] \\
$\psitrial^s$ & snapshot-based trial wave function in the channel $s$~\eqref{eq:trial_basis} applied to the KVP functional~\eqref{eq:kvp_functional} \\
$L^{s s'}_E$ & a generic scattering matrix at energy $E$ \\
$\genkvp^{ss'}[\psitrial]$ & a functional whose stationary point is an approximation of the generic $L$-matrix; i.e., $\genkvp[\psitrial + \delta\psitrial] = L^{s s'}_E+ \mathcal{O} (\delta L^2)$ \\
$\weights_i$ & to-be-determined coefficient of the $i\mathrm{th}$ snapshot in the trial wave function with $\sum_i \weights_i = 1$ \\
$\dU^{ss'}_{ij}(\param)$ & $\nbasis \times \nbasis$ kernel matrix defined in Eq.~\eqref{eq:delta_U_general}
\end{tabular}
\end{ruledtabular}
\end{table}
%%%%%%%%%%%%%%%%%%%%%%%%%%%%%%%%%%%%%%%%%%%%%%%%%%%%%%%%

Our goal is to emulate the partial-wave Schr\"odinger equation for $NN$ scattering at the center-of-mass energy $E > 0$
\begin{align} \label{eq:schrodinger}
    \widehat H(\param) \ket{\psi^{s}} \equiv \big[ \widehat T + \widehat V (\param) \big] \ket{\psi^{s}} = E  \ket{\psi^{s}},
\end{align}
where the vector $\param$ is composed of parameters used by the theoretical model to match results with experimental observations (e.g., the LECs of \chiEFT). Building our snapshot-based MOR emulator begins by writing Eq.~\eqref{eq:schrodinger} in integral form. Here, we choose the general (constrained\footnote{For a description of constrained and unconstrained emulators see Ref.~\cite{Drischler:2022ipa}}) KVP, which is based on the functional~\cite{Lucchese:1989zz,Drischler:2021qoy}
\begin{align} \label{eq:kvp_functional}
    \genkvp^{ss'}[\psitrial] 
    = \widetilde L^{s s'}_E - \frac{2 \mu \kzero}{\det \umatrix} \braket{\psitrial^{s} | \widehat H(\param) - E | \psitrial^{s'}},
\end{align}
where $\psitrial$ is a trial scattering wave function,
$\widetilde L^{s s'}_E$ is a generic trial scattering matrix, $\umatrix$ is a non-singular matrix~\cite{Lucchese:1989zz,Drischler:2021qoy} used to parametrize the asymptotic boundary condition associated with $\widetilde L^{s s'}_E$ (see Appendix~\ref{sec:anomalies}), and
$\kzero = \sqrt{2\mu E}$ is the on-shell energy with $\mu$ being the reduced mass.%
\footnote{Throughout this paper we use boldface symbols to indicate vectors in parameter-space, arrows to indicate vectors in snapshot-space, natural units in which $\hbar = c = 1$, and follow the conventions for scattering matrices in Refs.~\cite{taylor2006scattering,Morrison:2007}.} More details can be found in Ref.~\cite{Drischler:2021qoy} and Appendix~\ref{sec:anomalies}. Table~\ref{tab:notation} summarizes the notation we use in this work. Note that we adopt the convention that the wave functions in a bra symbol $\bra{\cdot}$ in bra-ket notation are not complex conjugated [e.g., $\bra{\psitrial^s}$ in Eq.~\eqref{eq:kvp_functional}]~\cite{10.1143/PTPS.62.236,Furnstahl:2020abp,Drischler:2021qoy}.

In Eq.~\eqref{eq:kvp_functional}, the superscripts $s$ and $s'$ index the coupled channels (e.g., ${^3}S_1$ and ${^3}D_1$); for the uncoupled case this reduces to a single equation with $s' = s$. Each combination of $(s',s)$ will have their own, distinct emulator in our formulation. As an example, for a coupled-channel \npr\ interaction in Eq.~\eqref{eq:kvp_functional}, the $(s',s)$ pair could be one of {${^3}S_1$--${^3}S_1$, ${^3}S_1$--${^3}D_1$, ${^3}D_1$--${^3}S_1$, or ${^3}D_1$--${^3}D_1$}, and for an uncoupled channel $s' = s$ could be ${^1}S_0$.
We use the \npr\ spin-triplet coupled channels as an exemplary case, but the general emulation procedure applies to general channel coupling (including spin-singlet spin-triplet \npr\ coupling~\cite{Stoks:1990us}).

The functional~\eqref{eq:kvp_functional} yields $\genkvp^{ss'}[\psitrial] = L^{s s'}_E$ when $\psitrial$ is the exact wave function, and provides a stationary approximation otherwise: $\genkvp^{ss'}[\psi + \delta\psi] = L^{s s'}_E + \mathcal{O}(\delta L^2)$. Rather than finding a wave function $\ket{\psi}$ that satisfies Eq.~\eqref{eq:schrodinger}, our task has now changed to finding a wave function that makes Eq.~\eqref{eq:kvp_functional} stationary for a given choice of $E$.

The key to creating an efficient PMOR emulator from Eq.~\eqref{eq:kvp_functional} is to use a snapshot trial wave function,
\begin{align} \label{eq:trial_basis}
    \ket{\psitrial^{s}}
    \equiv \sum_{i=1}^{\nbasis} \weights_i \ket{\psi_i^{s}},
\end{align}
where $\nbasis$ is the number of parameter vectors $\{ \param_i \}_{i=1}^{\nbasis}$ in the training set and $\{\ket{\psi_i^{s}}\}_{i=1}^{\nbasis}$ the associated high-fidelity solutions to Eq.~\eqref{eq:schrodinger}, obtained by solving the LS equation directly (see also Sec.~\ref{sec:results}).
These solutions are determined once in the offline stage. The to-be-determined basis coefficients $\vec{\weights}$ will \emph{not} be the same for all the channels, resulting in independent emulators for each $(s',s)$ pair (see Appendix~\ref{sec:coupled_channel_details} for more details). 
For the \npr\ spin-triplet coupled channels, this will result in three distinct variational principles being enforced: one for each of angular momentum $s' = s = j \pm 1$ and one for the off-diagonal component. The other off-diagonal component can be inferred through the unitarity of the $S$ matrix.%
\footnote{For (complex-valued) optical potentials with two coupled channels, one has four (instead of three) distinct variational principles because the $S$ matrix is not unitary.}

Upon inserting the snapshot trial wave function~\eqref{eq:trial_basis} into the functional~\eqref{eq:kvp_functional}, the functional takes the form~\cite{Furnstahl:2020abp} 
\begin{align} \label{eq:kvp_trial_functional}
    \genkvp^{ss'}  [\vec{\weights} \,]
    = \weights_i L^{s s'}_{E,i} - \frac{1}{2} \weights_i \dU_{ij}^{ss'} \weights_j,
\end{align}
with the symmetric matrix
\begin{align} \label{eq:delta_U_general}
    \dU^{ss'}_{ij}(\param) 
    & \equiv 
    \frac{2 \mu \kzero}{\det \umatrix}  \bigl[
    \braket{\psi_i^{s} | \widehat H(\param) - E | \psi_j^{s'}}
     + (i \leftrightarrow j) \bigr] \notag \\
    & = 
    \frac{2 \mu \kzero}{\det \umatrix} \bigl[
    \braket{\psi_i^{s} | \widehat V(\param) - \widehat V_j | \psi_j^{s'}}
   + (i \leftrightarrow j) \bigr],
\end{align}
where, as in Eq.~\eqref{eq:kvp_functional}, $s'$ and $s$ correspond to the entrance and exit channels. Equation~\eqref{eq:kvp_trial_functional} is a stationary approximation to the generic $L$ matrix at one energy, hence we build independent emulators for each value of an energy grid.
Equation~\eqref{eq:delta_U_general} is obtained~\cite{Furnstahl:2020abp} by adding and subtracting $\widehat V_i \equiv \widehat V(\param_i)$ and $\widehat V_j \equiv \widehat V(\param_j)$ and applying  Eq.~\eqref{eq:schrodinger}. In this form, the constant terms in the potentials, such as a long-range Coulomb interaction (assuming the fine-structure constant is not varied), will cancel, and the matrix elements will only involve short-range physics.

Emulating the scattering wave function [via Eq.~\eqref{eq:trial_basis}], and hence $L^{s s'}_E \approx \genkvp^{ss'}[\psitrial]$ [via Eq.~\eqref{eq:kvp_trial_functional}], has now been reduced to choosing an appropriate training set $\{ \param_i \}$ and then determining the values of $\weights_i$ that make Eq.~\eqref{eq:kvp_trial_functional} stationary under the constraint that $\sum_i \weights_i = 1$. The latter is a consequence of maintaining a consistent asymptotic normalization for the scattering wave functions in Eq.~\eqref{eq:trial_basis} as required by the constrained KVP~\cite{Furnstahl:2020abp, Drischler:2022ipa}. A numerically robust solution can be found by introducing a Lagrange multiplier $\lambda$, and solving the matrix equation~\cite{Drischler:2021qoy}
\begin{align} \label{eq:coeff_solution}
    \begin{pmatrix}
        \dU^{ss'} & \vec{1}\,{} \\ 
        \vec{1} \, {}^\intercal & 0\,{}
    \end{pmatrix}
    \begin{pmatrix}
        \kvpweights \, {} \\ 
        \lagmult \, {}
    \end{pmatrix}
    =
    \begin{pmatrix}
        \vec{L}^{ss'}_E \, {} \\ 
        1
    \end{pmatrix},
\end{align}
where $\vec{1}$ is an $\nbasis \times 1$ vector of ones, $\vec{L}^{ss'}_E$ are the basis states used in the offline stage, and $\vec{\weights}_\star$ is a vector of coefficients of the trial wave function associated with the KVP's stationary approximation. Since Eq.~\eqref{eq:coeff_solution} is a linear system, it will be a highly computationally efficient emulator for scattering systems if the number $\nbasis$ of basis functions is much smaller than the size of the high-fidelity wave function $\psi$.

Thus far we have not specified whether the matrix elements $\dU^{ss'}_{ij}$ are to be calculated in coordinate space or momentum space. The only difference between these implementations is the way we obtain the basis functions $\psi_i$ used to construct the trial ansatz in Eq.~\eqref{eq:trial_basis}, and thus the manner in which $\dU^{ss'}$ is evaluated. To formulate a momentum-space wave function approach to MOR emulators for scattering, we initially solve for the $K$ matrix and relate $\psi$ to $K$ before using Eq.~\eqref{eq:delta_U_general}. The scattering wave function in momentum space takes the form~\cite{Haftel:1970zz}
\begin{align} \label{eq:momentum_space_wf_coupled}
    \psi^{s t} (k; \kzero) = \frac{1}{k^2} \delta(k - \kzero) \delta^{s t} + \frac{2}{\pi} \mathbb{P} \frac{K^{s t} (k, \kzero) / \kzero}{k^2 - k^2_0}\,,
\end{align}
which vanishes as $k \rightarrow \infty$, but is singular at $k = \kzero = \sqrt{2\mu E}$ (the superscripts used for the $K$ matrix in Eq.~\eqref{eq:momentum_space_wf_coupled} are opposite Ref.~\cite{Haftel:1970zz}). Here, $K^{st}$ is the reactance matrix (or just the $K$~matrix), $\kzero$ the on-shell energy, $\mathbb{P}$ the Cauchy principal value, and the labeling $st$ indicates the partial-wave or reaction channels. One can also write Eq.~\eqref{eq:delta_U_general} in the momentum-space representation by inserting complete sets of states,%
\footnote{For example, for \npr\ scattering as in Sec.~\ref{sec:results}, the complete set of states are relative-momentum partial-wave states with orbital angular momentum and spin coupled to total $J$ and $M_J$.}
resulting in
\begin{align} \label{eq:delta_U_mom_representation}
    \dU^{ss'}_{ij}(\param) 
    = 
    \iint^{\infty}_0 \dd{k} \dd{p} k^2 p^2
    \bigl[
    &\psi_i^{ts}(k) V^{tt'}_{\param, j} (k,p) \psi_j^{t's'}(p) \notag \\
    \quad \quad
    &+ (i \leftrightarrow j)
    \bigr]
\end{align}
with
\begin{align} \label{eq:potential_definition}
    V^{tt'}_{\param, j} (k,p) \equiv \frac{2 \mu \kzero}{\det \umatrix} 
    \big[ V^{tt'}(k,p; \param) - V^{tt'}_j (k,p) \big],
\end{align}
where $t$ and $t'$ are summed over the available channels and the dependence of $\psi$ on $k_0$ is left implicit. Moving forward, we will drop the channel superscripts on $\dU$.

This is the general form of the momentum-space $\dU$ matrix. Note the ordering of the channel indices $(t,s)$ in the left-hand wave function in Eq.~\eqref{eq:delta_U_mom_representation}, which follows from $\psi^{ts}(k) \equiv \braket{kt | \psi^s}$ and the convention that $\bra{\psi} = \ket{\psi}^\trans$ (without a complex conjugate), so that $\psi^{ts}(k) = \braket{\psi^s | kt}$. Thus, if $\psi$ has outgoing ($\psi^{(+)}$), incoming ($\psi^{(-)}$), or standing wave ($\psi^{(0)}$) boundary conditions, then the same version of $\psi^{(x)}$ is used for both $\psi(k)$ and $\psi(p)$ in Eq.~\eqref{eq:delta_U_mom_representation}. No modification of Eq.~\eqref{eq:delta_U_mom_representation} is needed in the case of optical potentials, where again the left-hand wave function is \emph{not} conjugated relative to the right-hand wave function. For more details on how to build the general KVP emulator we refer the reader to Appendix~\ref{sec:emulator_details}. Different boundary conditions will be used below to mitigate Kohn anomalies (see Sec.~\ref{sec:results_partial_waves}).

The efficient evaluation of $\dU$ across a range of $\param$ values is critical to the applicability of the emulator. If the Hamiltonian operators have an affine (i.e., factorizable) parameter dependence, denoted as
\begin{equation} \label{eq:def_affine}
    \widehat H(\param) = \sum_n h_n(\param) \widehat H_n,
\end{equation}
then matrix elements of the $H_n$ operators in a given basis only need to be calculated once in the offline stage rather than for every parameter set $\param_i$. Chiral $NN$ interactions have the form of Eq.~\eqref{eq:def_affine} and, when varying only the contact LECs, can even be cast into the form%
\footnote{Note that $h_n(\param)$ would include higher-order polynomials when also emulating the pion-nucleon coupling $c_2$ (at \NNNLO) and axial coupling constant $g_A$ (already at LO). Nevertheless, the Hamiltonian remains affine and thus the emulators discussed here are directly applicable.}
\begin{align} \label{eq:linear_params_pot}
    \widehat V(\param) 
    = 
    \widehat V^0 + \param \cdot \boldsymbol{\widehat V}^1,
\end{align}
so that Eq.~\eqref{eq:delta_U_general} can then be written as
\begin{align} \label{eq:delta_U_linear}
    \dU (\param)
    = 
    \dU^{0} 
    + \param \cdot \Delta \boldsymbol{\widetilde {U}}^{1}.
\end{align}
The matrices $\widehat V^0$ and $\dU^{0}$ and vectors of matrices $\boldsymbol{\widehat V}^1$ and $\Delta \boldsymbol{\widetilde {U}}^{1}$, 
can now be pre-calculated during the emulator's offline stage, allowing for considerable speed-up factors in the online stage where the value of $\dU (\param)$ at any new parameter value is efficiently constructed.

%%%%%%%%%%%%%%%%%%%%%%%%%%%%%%%%%%%%%%%%%%%%%%%%%%%%%%%%
\section{Results} \label{sec:results}
%%%%%%%%%%%%%%%%%%%%%%%%%%%%%%%%%%%%%%%%%%%%%%%%%%%%%%%%
In this section, we apply the KVP momentum-space emulator to calculate \npr\ scattering observables. We use the Reinert \etal\ semilocal momentum-space (SMS) regularized chiral potential at N$^4$LO$+$ with the momentum cutoff $\Lambda = 450\MeV$~\cite{Reinert:2017usi}, which is a state-of-the-art chiral $NN$ interaction. The parameters $\param$ are composed of the $NN$ contact LECs contributing to this potential.

\subsection{Emulator overview} \label{sec:overview}
The snapshots used in the offline stage are the scattering solutions given by Eq.~\eqref{eq:momentum_space_wf_coupled}. The $K$ matrices used to calculate the second term in Eq.~\eqref{eq:momentum_space_wf_coupled} are obtained from numerically solving the LS equation. The LS equation is reduced to a set of linear equations by approximating the integral as a sum over $N$ quadrature points obtained from Gauss--Legendre rules with corresponding weights (see Refs.~\cite{Haftel:1970zz, Landau:1996}). If the potential was calculated merely on the quadrature points, without appending the on-shell values, interpolation must be performed to obtain the (half-)on-shell potential so that one can (1) account for the singularity of the Green's function when solving the LS equation~\cite{Landau:1996}, and (2) integrate the delta distribution in Eq.~\eqref{eq:momentum_space_wf_coupled}. 

To generate the figures in this paper, we use a three-segment compound Gauss-Legendre quadrature mesh with a total of $N=80$ momentum points. Half of the points are placed in the first segment ($0$--$3\fmi$) and the other half split between the second ($3$--$6\fmi$) and third segment ($6$--$\infty\fmi$). The total number of points was informed by prior experience solving the Lippmann-Schwinger equation for similar potentials. Furthermore, tests on a separable potential indicated that the relative error between the exact and simulator solutions begins to saturate around that point. 
We emphasize that the momentum mesh is only relevant in the offline stage since this directly affects the precision of the high-fidelity solutions from the simulator used to build the basis, and therefore the precision of the emulator with respect to the \emph{exact solution} (solution as $N \rightarrow \infty$). The choice of mesh does not affect the emulator's implementation in the online stage, e.g., it does not affect the size of the $\dU$ matrix in Eq.~\eqref{eq:delta_U_general}. For the observables, we use a laboratory energy range of $0.1$ to $350 \MeV$ with 350 points. For the partial waves plots, we use a fine energy mesh of 3500 points over the same energy range.

When performing the KVP emulation, we calculate Eq.~\eqref{eq:delta_U_general} two different ways. The first is by inserting Eq.~\eqref{eq:momentum_space_wf_coupled} into Eq.~\eqref{eq:delta_U_general} and analytically integrating the $\delta$ distribution, which corresponds to appending the exact on-shell value of the potential. The remaining integrals are then solved numerically (see Appendix~\ref{sec:emulator_details}). We refer to this method as the standard method. The second is based on the global Gl{\"o}ckle spline interpolation~\cite{GlockleInterpolation1982}, which belongs to the family of quasispline methods that perform the mapping
\begin{equation}
    \sum_k f(k) \mathcal{S}_k(\kzero) \approx f(\kzero),
\end{equation}
for smooth functions $f(k)$ sampled on a grid $k$ that encompasses $\kzero$ using the cubic spline polynomials $\mathcal{S}_k (\kzero)$ constructed in Ref.~\cite{GlockleInterpolation1982}. This allows us to calculate $\mathcal{S}_k (\kzero)$ once in the offline stage and save the result for the online stage since it has no dependence on $f(k)$ itself. Using this method, we interpolate the solutions to the integrals that appear in Eq.~\eqref{eq:delta_U_general} (i.e., $\kzero$ does not need to be appended to the mesh as opposed to the standard method), thus decreasing the computational cost needed in the offline stage significantly at the expense of accuracy. We compare the KVP emulator results using the Gl{\"o}ckle and standard method and compare those results to the NVP emulator described in Ref.~\cite{Melendez:2021lyq}.

To reduce numerical errors in both the simulator and emulator, we compute snapshots $\{K_i\}$ of the LS equation using non-interpolated potentials for partial waves that have a LEC-dependence and interpolated potentials for LEC-independent partial waves. When referring to interpolated potentials, we mean calculating the potential using only the momentum mesh and then using an interpolation method (such as the bivariate Gl{\"o}ckle spline method) to interpolate the potential to $\kzero$. By noninterpolated, we mean that each $\kzero$ is appended to the momentum mesh and the potential evaluated at these points, which improves the accuracy of our potentials compared to interpolating the potential to $\kzero$. We chose to use non-interpolated potentials for the LEC-dependent partial waves since these are the only ones used to calculate Eq.~\eqref{eq:delta_U_general} in the offline phase. The same LEC-independent partial waves are employed by the simulator and emulator. All potentials used for the emulators and simulator are precalculated for efficiency.

The simulator used in this paper numerically solves the LS equation for each partial wave. The accuracy of our simulator was tested by comparing the simulator results to the analytical solution of a Gaussian separable potential, producing relative errors of $\approx 10^{-7}$ or better. Additionally, the simulator's speed was roughly $4\times$ slower when we doubled the mesh size from $N=80$ to $N=160$ quadrature points (by doubling the points in each segment).

The accuracy of emulated observables depends on the size of the basis (see Sec.~\ref{sec:results_observables}); here we use a basis size $\nbasis = 2 \nlecs$, where $\nlecs$ is the number of LECs associated with a given partial wave channel. The training points $\param_i$ are randomly sampled within an interval of $[-5, 5]$ using a Latin-hypercube for each partial wave, with the fitted coupling constants and appropriate units given in Ref.~\cite{Reinert:2017usi}.

The matrix $\dU$ is increasingly ill-conditioned as the basis size $\nbasis$ increases.
One can reduce numerical noise by (1) adding a regularization parameter (``nugget'') to the diagonal elements of the near-singular matrix~\cite{Furnstahl:2020abp}, or (2) using a solver that performs some type of regularization. For the KVP emulator results in the figures, we use NumPy's least-squares solver \texttt{linalg.lstsq()}~\cite{harris2020array} with a cut-off ratio for small singular values of $10^{-10}$~\cite{Drischler:2021qoy}. For the NVP emulator, we add a nugget of $10^{-10}$ to the diagonal and use NumPy's \texttt{linalg.solve()}.

The general KVP functional may not always provide a (unique) stationary approximation, giving rise to spurious singularities known as Kohn (or Schwartz) anomalies~\cite{PhysRev.124.1468, nesbet1980variational}. The energies at which those anomalies occur depends on the training parameters $\param$ used in the offline stage and the evaluation set used in the online stage.
Reference~\cite{Drischler:2021qoy} proposed detecting and mitigating these numerical instabilities by considering an array of KVPs with different boundary conditions (i.e., scattering matrices) within a partial wave and using the emulator solutions to obtain an estimated $S$ matrix by a weighted sum of averages (see also Refs.~\cite{Lucchese:1989zz,Viviani:2001sy}).

For our KVP emulator, the mitigation process involves first calculating Eq.~\eqref{eq:delta_U_general} using the $K$ matrix boundary condition. Once we have calculated $\dU$, the terms in Eq.~\eqref{eq:kvp_trial_functional} are rescaled to match the boundary conditions we want to emulate (here, $L=K$, $K^{-1}$, and $T$). The anomalies are then detected by applying a consistency check to the (independent) emulated solutions of the different boundary conditions. The emulator solutions that do not pass this check are discarded while those that pass are averaged to obtain an anomaly-free scattering matrix (here, the $S$ matrix). All KVP emulator results in this paper are shown with anomaly mitigation unless otherwise stated. So far, such a mitigation protocol has not been implemented for the NVP emulator. However, one approach would be to use multiple emulators based on different variational principles~\cite{Drischler:2022ipa} instead of multiple boundary conditions. See Appendix~\ref{sec:anomalies} for our implementation and Ref.~\cite{Drischler:2021qoy} for more information on emulation with arbitrary boundary conditions and ways to mitigate Kohn anomalies.

%%%%%%%%%%%%%%%%%%%%%%%%%%%%%%%%%%%%%%%%%%%%%%%%%%%%%%%%
\subsection{Emulation of phase shifts} \label{sec:results_partial_waves}
%%%%%%%%%%%%%%%%%%%%%%%%%%%%%%%%%%%%%%%%%%%%%%%%%%%%%%%%

\begin{figure}[tb]
    \centering
    \includegraphics
    {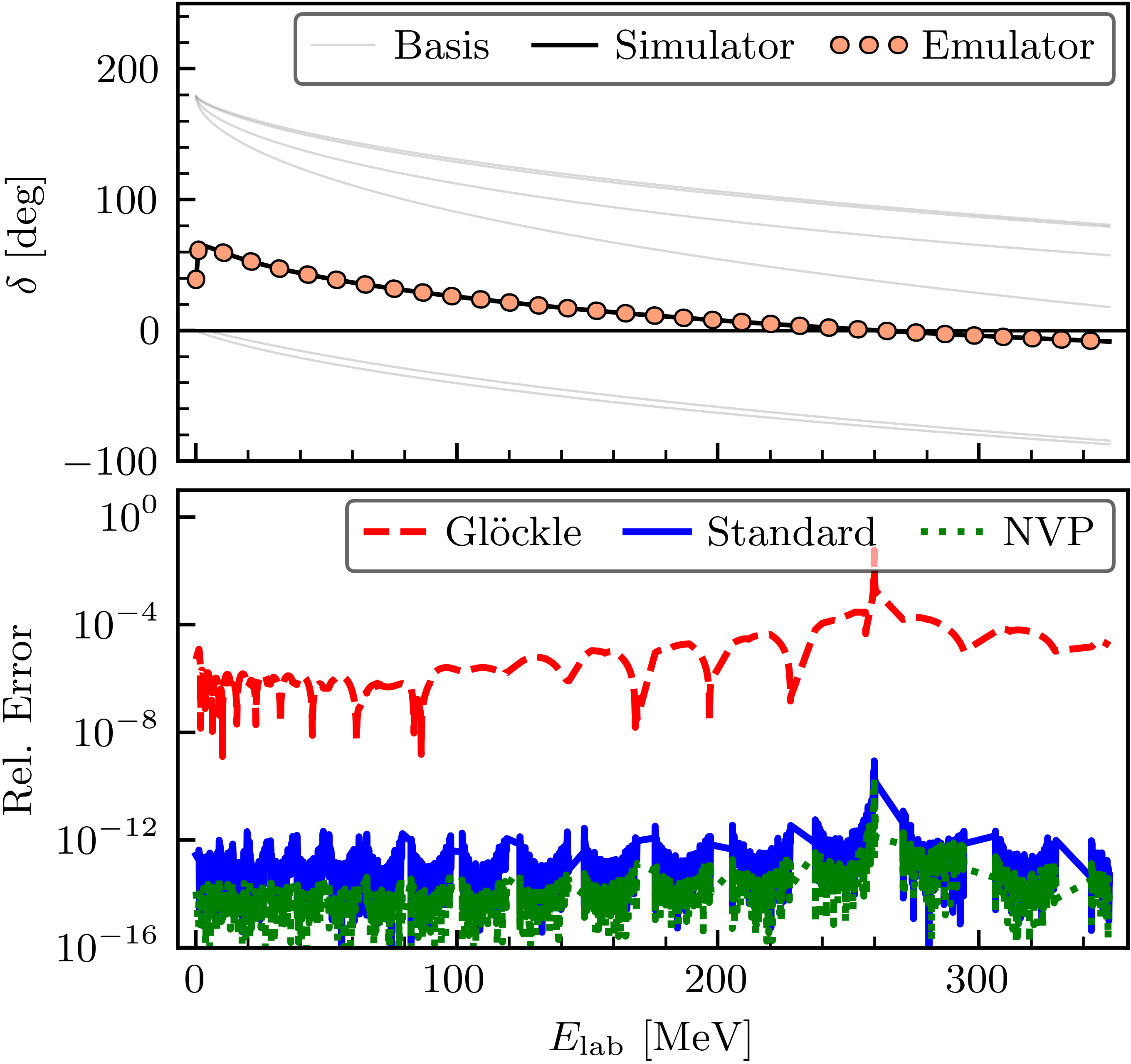}
    \caption{%
        Simulated (black solid line) and KVP emulated standard method (orange dots) ${^1}S_0$ phase shifts for the N$^4$LO$+$ SMS potential with $\Lambda = 450 \MeV$ (top panel).
        The bottom panel shows the relative errors between the simulated and emulated phase shifts for the Gl{\"o}ckle method (red dashed line), standard method (blue solid line), and NVP emulator (green dotted line), respectively.
        The spike at $\Elab \approx 270 \MeV$ is due to the phase shift crossing zero.
    }
    \label{fig:np_1S0}
\end{figure}

We first apply the emulators to the uncoupled ${^1}S_0$ channel using Eq.~\eqref{eq:delta_U_general} to calculate $\dU$ (see Appendix~\ref{sec:emulator_details} for explicit expressions). At N$^4$LO$+$, this channel depends on $\nlecs = 3$ non-redundant LECs~\cite{Reinert:2017usi}, and thus we choose our basis to be composed of $\nbasis = 6$ training points. Figure~\ref{fig:np_1S0} shows the phase shifts calculated using our simulator (black line) and the KVP emulator standard method prediction (orange dots) as a function of the laboratory energy in the top panel. The phase shifts associated with the training points are depicted by the light gray lines. In addition, the bottom panel shows the relative errors
\begin{align}
    \text{Rel.\ Error} = 2 \left|
    \frac{\text{Simulator} - \text{Emulator}}{\text{Simulator} + \text{Emulator}}
    \right|
\end{align}
between the simulated and emulated phase shifts for the Gl{\"o}ckle method (red dashed line),  standard method (blue solid line), and NVP emulator (blue dotted line).

\begin{figure*}[tb]
    \centering
    \includegraphics
    {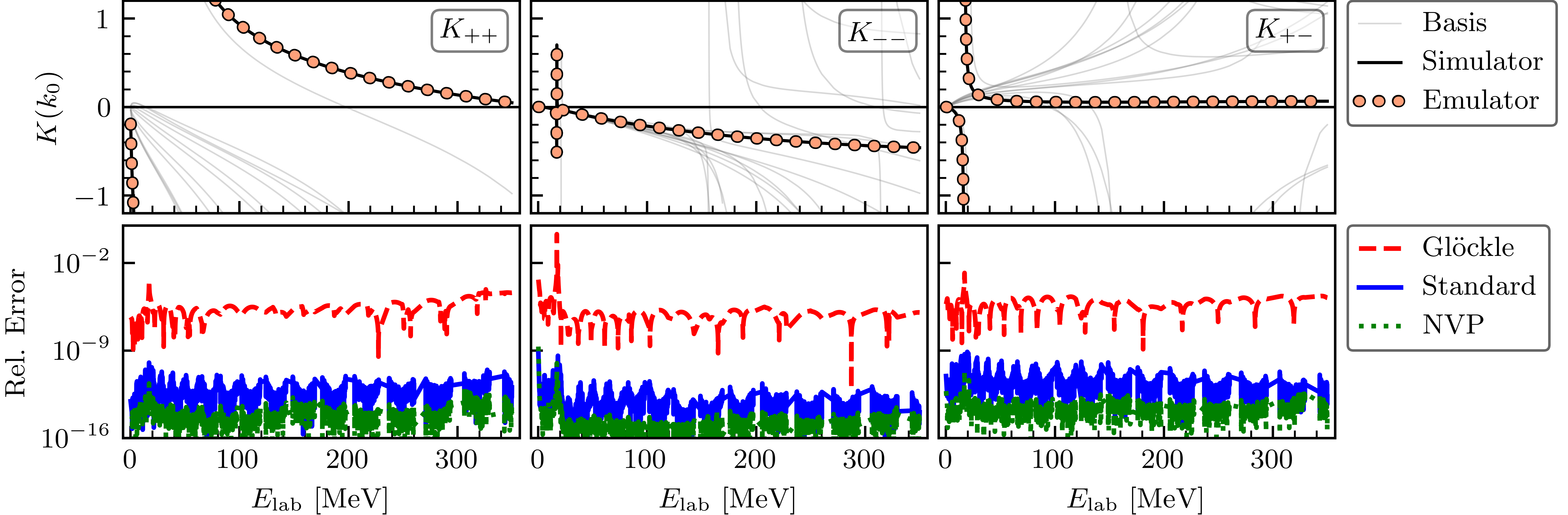}
    \caption{%
        As in Fig.~\ref{fig:np_1S0}, but for the
        on-shell $K$ matrix in 
        the coupled ${^3}S_1$--${^3}D_1$ as a function of the laboratory energy.
        From left to right: pure $D$ wave, pure $S$ wave, and mixed $S$-$D$ wave component.
    }
    \label{fig:np_3S1_3D1_coupled}
\end{figure*}

We find that our KVP emulator accurately reproduces the high-fidelity phase shifts over a large energy range for both methods, but the standard method is much more accurate than the Gl{\"o}ckle method. On average, the relative error for the Gl{\"o}ckle method is on the order of $\approx 10^{-6}$--$10^{-5}$, while the standard method has a relative error on the order of $\approx 10^{-12}$ for the same basis size. The NVP emulator's relative error is similar to the KVP standard method, with an error of $\approx 10^{-13}$.

We now turn to the coupled ${^3}S_1$--${^3}D_1$ channel. This channel depends on $\nlecs = 6$ non-redundant LECs~\cite{Reinert:2017usi} at N$^4$LO$+$, which means that our basis will be composed of $\nbasis = 12$ training points. Figure~\ref{fig:np_3S1_3D1_coupled} shows the on-shell $K$ matrix for the simulator calculation (black lines) and KVP emulator prediction (orange dots) as a function of the laboratory energy for each different partial-wave component. The errors are similar to the ${^1}S_0$ channel, with the standard method being much more accurate than the Gl{\"o}ckle method, and the NVP emulator's relative error being slightly better than the standard method. In all cases, we see a spike in the relative error at $\Elab\approx 20 \MeV$ where the $K$ matrix is singular.

The small spikes seen in the standard method error are \emph{not} Kohn anomalies, but can be attributed to a numerical instability of the principal value integral in the LS equation. These spikes are mesh-dependent and appear when a $\kzero$ value is close to a momentum mesh point, thus causing the denominator of the Green's function to approach zero faster than the numerator. A way to decrease the relative error produced by these spikes is to not allow the $\kzero$ values to be close to momentum mesh points by moving energies that are close to any momentum mesh point until the relative distance is greater than some threshold value; e.g., $\varepsilon \gtrsim 10^{-2} \MeV$ (see Appendix~\ref{sec:additional_resuls} for details). The oscillations that appear in the Gl{\"o}ckle method's relative errors plots are potential-dependent, and increase in number, but decrease in separation, when increasing the mesh size.

Overall, the emulators accurately predict the partial waves for the uncoupled ${^1}S_0$ and coupled ${^3}S_1$--${^3}D_1$ channels. When comparing the Gl{\"o}ckle method emulation with the standard method, we see that the relative error for the standard method is much less than the Gl{\"o}ckle method. For both partial waves shown, the NVP emulator is the one that most accurately reproduces its high-fidelity solution. Results for the other channels are similar to the ones presented here, with the only difference being that the relative error decreases as $\nlecs$ gets smaller. This can be further explored with the Jupyter notebooks provided~\cite{BUQEYEsoftware}.

%%%%%%%%%%%%%%%%%%%%%%%%%%%%%%%%%%%%%%%%%%%%%%%%%%%%%%%%
\subsection{Emulation of scattering observables} \label{sec:results_observables}
%%%%%%%%%%%%%%%%%%%%%%%%%%%%%%%%%%%%%%%%%%%%%%%%%%%%%%%%
Next, we examine the performance of the emulator for nuclear observables. As a demonstration, we use the SMS regularized chiral potential at N$^4$LO$+$ for \npr\ scattering with partial waves having total momentum quantum numbers  $j \leqslant j_{\mathrm{max}} = 20$. Overall, there are a total of $25$ parameters in $\param$ that are being sampled using a Latin-hypercube design. As previously mentioned, the basis size is chosen as $\nbasis = 2 \nlecs$, where $\nlecs$ is the number of LECs associated with the specific partial wave, for a total of 50 training points. Since these parameters are only present in the channels $j \leqslant 4$, the emulator only needs to be trained over these channels. The remaining channels do not change as the parameters are varied, therefore, they do not undergo a training process and need to be calculated only once by solving the LS equation directly.

\begin{figure}[tbhp!]
    \centering
    \includegraphics
    {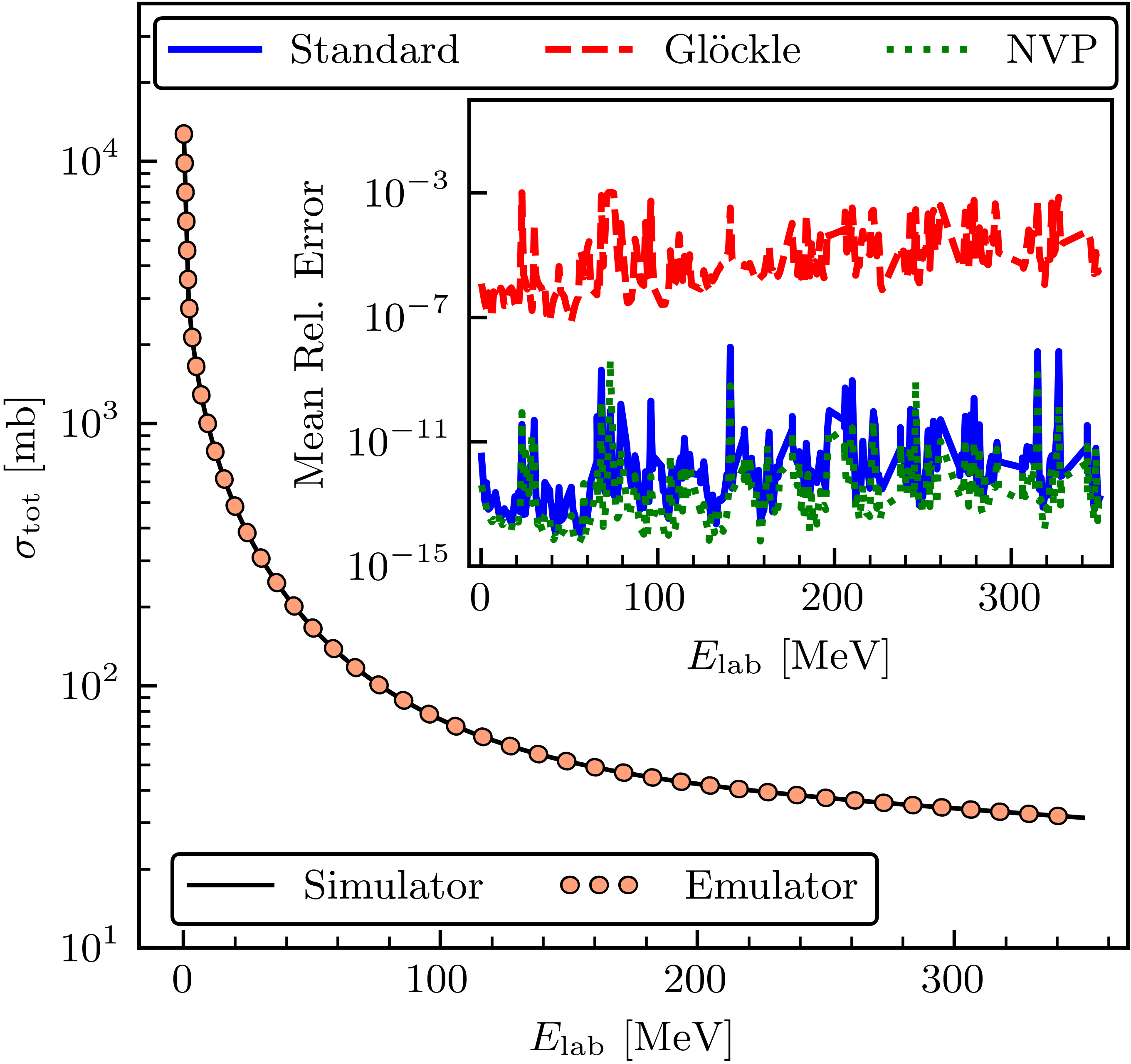}
    \caption{
        Simulated (black solid line) and emulated (orange dots) \npr\ cross section with $j_{\mathrm{max}} = 20$ for the N$^4$LO$+$ SMS potential with $\Lambda = 450 \MeV$ as a function of the laboratory energy. The inset shows the relative mean errors between the emulator and the simulator using the Gl{\"o}ckle, standard method, and NVP emulator for 500 different sets of \npr\ LECs obtained from Latin-hypercube sampling. See the main text for details.
     }%
    \label{fig:cross_section}%
\end{figure}

The emulation of observables is carried out by combining multiple emulators across different partial-wave channels. The total \npr\ cross section can be calculated using 
\begin{equation}
\sigma_{\text{tot}}(\kzero) 
= 
\frac{\pi}{2 \kzero^2} \sum^{j_\mathrm{max}}_{j=0} 
  (2j + 1) \, \mathrm{Re} \{\mathrm{Tr} [S_j(\kzero) - \mathds{1}_4] \},
\end{equation}
where $S_j = \mathds{1}_4 - 2i (\mathds{1} - i K_j)^{-1} K_j$ is the $S$ matrix, $K_j$ is the predicted on-shell $K$ matrix, and $\mathrm{Tr}[\cdot]$ denotes the trace. Both $S_j$ and $K_j$ are $4\times 4$ matrices that contain both the triplet-triplet and the singlet-triplet channels.

Figure~\ref{fig:cross_section} shows the simulator and emulator prediction for the total \npr\ cross section, which are calculated using the fit values for the LECs determined in Ref.~\cite{Reinert:2017usi}. The inset in Fig.~\ref{fig:cross_section} depicts the mean relative errors for all three emulators when randomly sampling 500 different combinations of \npr\ LECs (chosen within the same range as the training points), using these to calculate the emulated and simulated total cross section, and comparing the results. On average, the relative errors for all three emulators are similar to those for the partial-wave calculations discussed in Sec.~\ref{sec:results_partial_waves}. Although the mean relative errors for the standard method and NVP emulators are very similar, the NVP emulator seems to be the one that most accurately reproduces its simulator.

As mentioned in Sec.~\ref{sec:overview} and following Ref.~\cite{Drischler:2021qoy}, the Kohn anomalies found in the calculation were mitigated by emulating with different boundary conditions and building the estimated $S$ matrix. Figure~\ref{fig:cross_section_spikes} in Appendix~\ref{sec:additional_resuls} shows a total cross section emulation result with one boundary condition, hence no anomaly mitigation. From the figure, we see that anomalies contribute to the standard method mean relative error at higher energies with a magnitude of approximately $10^{-3}$. These spikes are reduced to approximately $10^{-9}$ with mitigation. The Gl\"{o}ckle method result exhibits anomaly contributions of order $10^{-3}$ at lower energies, which get reduced to approximately $10^{-5}$--$10^{-7}$ with mitigation. For additional information, see the discussion in Appendix~\ref{sec:additional_resuls}. Although the NVP emulator is subject to anomalies, they are not evident in the figures shown in this section, even though no mitigation strategy was applied. An example of noticeable anomaly contributions as large as $10^{-3}$ in the NVP emulation are seen in Fig.~\ref{fig:cross_section_500MeV} in Appendix~\ref{sec:additional_resuls}.

The remaining spikes in Fig.~\ref{fig:cross_section} (e.g., at $\Elab \approx 140 \MeV$) can be traced back to singularities in the on-shell $K$ matrix for the ${^3}S_1$--${^3}D_1$ channel at those energies and are only seen for a few (specific) LEC values out of the 500 sampled (see also Fig.~\ref{fig:np_3S1_3D1_coupled}). The mesh-induced spikes seen in the standard method relative error were also reduced in magnitude by preventing the on-shell $\kzero$ value from being too close to a momentum mesh value (see Fig.~\ref{fig:1S0_spikes_test} for result comparisons).

We now turn our attention to spin-dependent observables for non-identical particles. A detailed description of $NN$ observables and their different conventions can be found in Refs.~\cite{Carlsson:2015vda, Stoks:1993tb,Bystricky:1976jr, lafrance:jpa-00208966, moravcsik:jpa-00210988, Stoks:1990us}. In general, one can write the spin observables in terms of Saclay parameters, which are complex functions of the center-of-mass energy and angle $\theta$. Here we only consider the differential cross section and analyzing power:
\begin{align} \label{eq:spin_obs}
    \dv{\sigma}{\Omega} = \, &\frac{1}{2} \Big[ |a|^2 + |b|^2 +|c|^2 +|d|^2 + |e|^2 + |f|^2 \Big] ,\\
    \dv{\sigma}{\Omega} A_y = \, &\mathrm{Re} (a^* \, e + b^* \, f),
\end{align}
where $d\sigma / d\Omega$ is the unpolarized differential cross section and $A_y$ the analyzing power (also known as $P_b$). For identical particles, one has $f=0$. More information on the description of the spin observables can be found in Refs.~\cite{Bystricky:1976jr, lafrance:jpa-00208966}; see also Appendix~\ref{sec:additional_resuls}, which contains our emulation results for more spin observables. The Saclay parameters can be obtained from the spin-scattering $M = M(\theta, \phi)$ matrix written in singlet-triplet space, 
\begin{align} \label{eq:M_scatting}
    M = 
    \begin{pmatrix}
          M_{11} & M_{10}e^{-i \phi} & M_{1-1}e^{-2i \phi} & M_{ST}e^{-i \phi} \\
          M_{01}e^{i \phi} & M_{00} & M_{0-1}e^{-i \phi} & 0 \\
          M_{-11}e^{2 i \phi} & M_{-10}e^{i \phi} & M_{-1-1} & M_{ST}e^{i \phi} \\
          M_{ST}e^{i \phi} & 0 & -M_{ST}e^{-i \phi} & M_{SS}
    \end{pmatrix}, 
\end{align}
where the subscripts $SS$ and $ST$ represent the singlet-singlet and singlet-triplet channel, respectively~\cite{Stoks:1993tb}. Equation~\eqref{eq:M_scatting} can be calculated using spherical harmonics and Clebsch-Gordan coefficients, and can be related to the Saclay parameters from the expressions:
\begin{align} \label{eq:saclay_params}
    a &= \frac{1}{2} (M_{11} + M_{00} - M_{1-1}) ,\\
    b &= \frac{1}{2} (M_{11} + M_{ss} - M_{1-1}) ,\\
    c &= \frac{1}{2} (M_{11} - M_{ss} - M_{1-1}) ,\\
    %d &= \frac{1}{2 \cos \theta} (-M_{11} + M_{00} + M_{1-1}) ,\\
    d &= -\frac{1}{\sqrt{2} \sin \theta} (M_{01} + M_{01}) ,\\
    e &= \frac{i}{2} (M_{10} - M_{01}) ,\\
    f &= -i \sqrt{2} M_{ST}.
\end{align}

\begin{figure}[tb]
    \centering
    \includegraphics
    {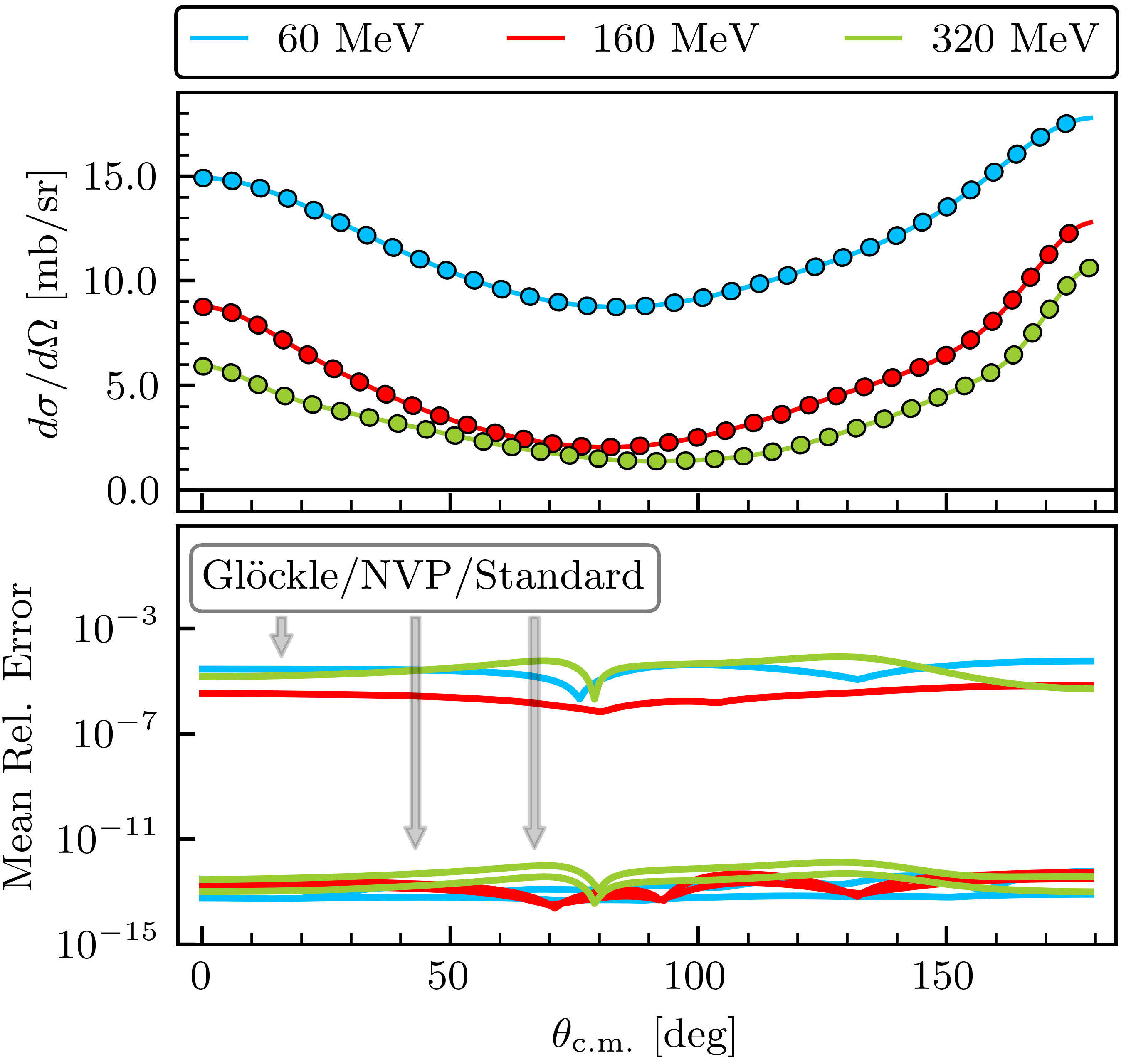}
    \caption{%
    Simulated (solid lines) and emulated (dots) unpolarized differential cross section for the N$^4$LO$+$ SMS potential with $\Lambda = 450 \MeV$ as a function of the center-of-mass angle at the three energies $60$, $160$, and $320 \MeV$ (top panel). The bottom panel shows the mean relative errors between the emulators and their respective simulators for 500 different sets of \npr\ LECs obtained from Latin-hypercube sampling. The colors for the relative mean errors correspond to the energies in the top panel. The gray arrows point from the label associated with the emulator to its error. See the main text for details.
    }
    \label{fig:differential_cs}
\end{figure}

\begin{figure}[tb]
    \centering
    \includegraphics
    {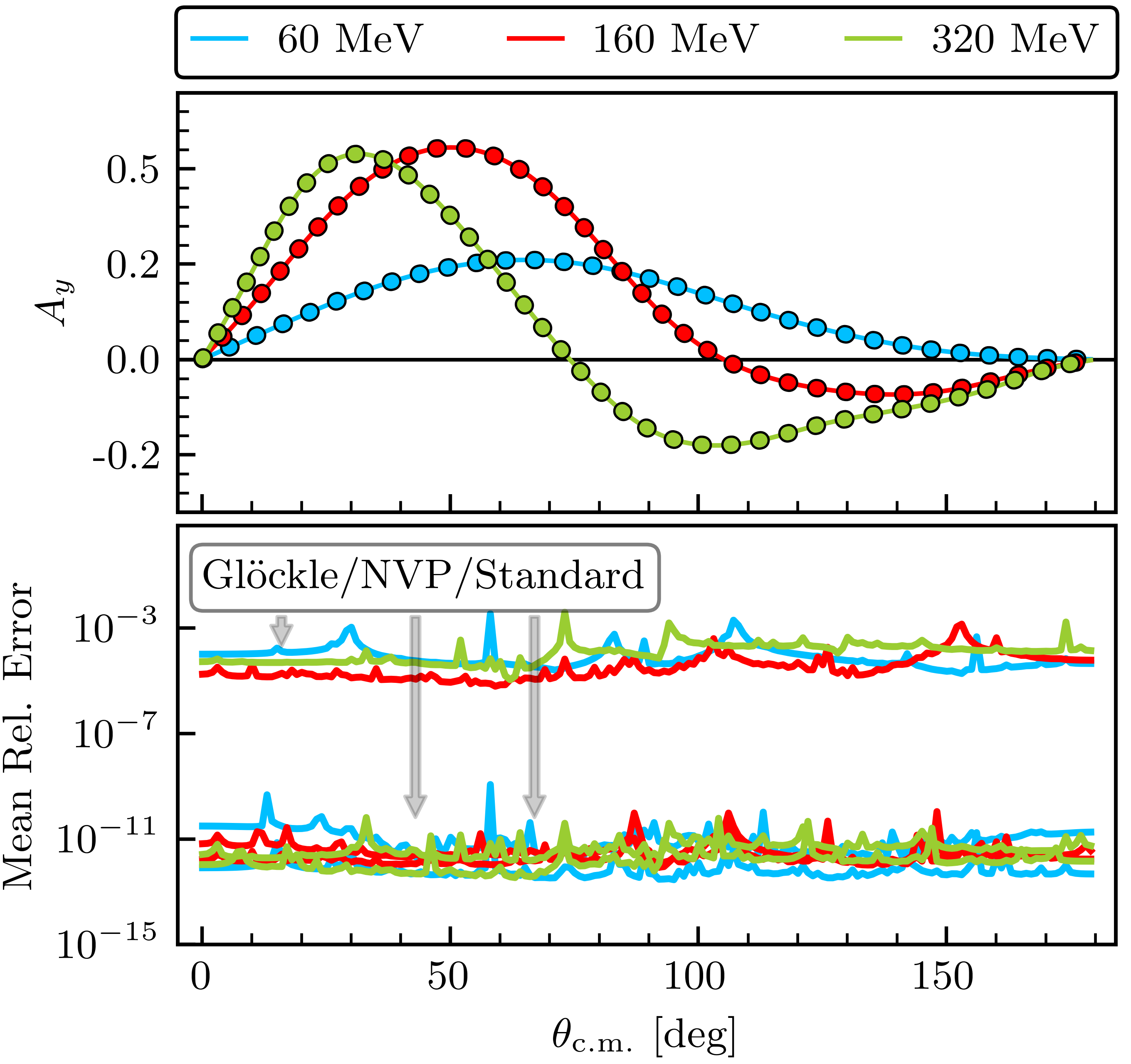}
    \caption{%
    As in Fig.~\ref{fig:differential_cs}, but for the analyzing power $A_{y}$ (also known as $P_b$). See the main text for details.
    }
    \label{fig:spin_correlation_amp}
\end{figure}

The emulation process is performed similarly to the one for the total cross section, where multiple trained emulators are combined across different partial-wave channels. Figures~\ref{fig:differential_cs} and~\ref{fig:spin_correlation_amp} show the simulator and emulator prediction for the differential cross section and analyzing power at three different energies calculated using the fit values for the LECs determined in Ref.~\cite{Reinert:2017usi}. The relative mean errors shown are obtained by randomly sampling 500 different combinations of \npr\ LECs (the same LECs used for the sampled relative error calculation in Fig.~\ref{fig:cross_section}) and comparing them against their respective simulator calculation. On average, the spin observables emulator has a relative mean error on the order of $\approx 10^{-5}$ when employing the Gl{\"o}ckle method and $\approx 10^{-14}$--$10^{-11}$ when using the standard method and NVP emulators, which are similar to the total cross section results. The results are similar to those obtained over the entire energy grid and for other observables (see Appendix.~\ref{sec:additional_resuls}).

Table~\ref{tab:spin_obs_results} details the angle-averaged relative errors between the simulator and KVP emulators (base-10 logarithm) for different spin observables with varying basis size at a variety of energies. As can be seen, when training the emulator with basis size $\nbasis = \nlecs$ both the standard and Gl{\"o}ckle method emulators have large relative errors of roughly $10^{-1}$ when compared to the high-fidelity model calculation. However, if we increase the basis size by doubling the parameters used per partial-wave, $\nbasis = 2 \nlecs$, the relative mean errors are significantly smaller, roughly $10^{-12}$--$10^{-9}$ and $10^{-6}$--$10^{-3}$, respectively. According to Ref.~\cite{Perez:2013mwa}, the relative errors given by $\nbasis = 2 \nlecs$ are below experimental uncertainties. When increasing the basis size to $\nbasis = 4 \nlecs$, the mean errors have mostly saturated and the improvement in accuracy is insignificant compared to the basis size $\nbasis = 2 \nlecs$. Note that although only four energies are shown, these results are similar over the entire energy grid.

\begin{table*}[tb]
\renewcommand{\arraystretch}{1.2}
\caption{Comparison of the angle-averaged relative errors (base-10 logarithm) between high-fidelity model and emulator for various angular observables with different basis size for 500 sets of \npr\ LECs using the N$^4$LO$+$ SMS potential~\cite{Reinert:2017usi} with momentum cutoff $\Lambda = 450 \MeV$ (rounded to two significant figures). These results are similar over the entire energy mesh. Here, ``Std.'' refers to the standard method emulator. See the main text for details.}
\label{tab:spin_obs_results}
\begin{ruledtabular}
\addtolength{\tabcolsep}{-2.5pt}
\begin{tabular}{lccccccccccc}
    & 
    & \multicolumn{2}{c}{$\dd\sigma/\dd\Omega$}
    & \multicolumn{2}{c}{$D$}
    & \multicolumn{2}{c}{$A_y$} 
    & \multicolumn{2}{c}{$A_{yy}$}
    & \multicolumn{2}{c}{$A$} \\
    % \cline{3-12}
    \cmidrule(lr){3-4}
    \cmidrule(lr){5-6}
    \cmidrule(lr){7-8}
    \cmidrule(lr){9-10}
    \cmidrule(lr){11-12}
    Basis size
    & $E$ [MeV]
    & Std.\ & Gl{\"o}ckle
    & Std.\ & Gl{\"o}ckle
    & Std.\ & Gl{\"o}ckle 
    & Std.\ & Gl{\"o}ckle 
    & Std.\ & Gl{\"o}ckle \\ 
    % \clineB{1-12}{2}
    \midrule
    {\multirow{4}{*}{$\nbasis = \nlecs$}} 
    & \multicolumn{1}{ c }{5}
    & $-1.2$ & $-1.2$
    & $-0.93$ & $-0.93$
    & $-0.46$ & $-0.46$
    & $-0.72$ & $-0.72$
    & $-0.78$ & $-0.78$
     \\
    & \multicolumn{1}{ c }{100}
    & $-0.73$ & $-0.73$
    & $-0.47$ & $-0.47$
    & $-0.12$ & $-0.12$
    & $-0.20$ & $-0.20$
    & $-0.28$ & $-0.28$
     \\
    & \multicolumn{1}{ c }{200}
    & $-0.54$ & $-0.64$
    & $-0.30$ & $-0.30$
    & $-0.028$ & $-0.028$
    & $-0.035$ & $-0.035$
    & $-0.12$ & $-0.12$
     \\
    & \multicolumn{1}{ c }{300}
    & $-0.49$ & $-0.49$
    & $-0.24$ & $-0.24$
    & $-0.066$ & $-0.066$
    & $-0.037$ & $-0.037$
    & $-0.043$ & $-0.043$
     \\
    % \clineB{1-12}{2}
    \midrule
    {\multirow{4}{*}{$\nbasis = 2 \nlecs$} } 
    & \multicolumn{1}{ c }{5}
    & $-10$ & $-7.0$
    & $-8.8$ & $-6.1$
    & $-8.8$ & $-5.6$
    & $-8.5$ & $-5.8$
    & $-8.3$ & $-5.9$
     \\
    & \multicolumn{1}{ c }{100}
    & $-12$ & $-6.3$
    & $-11$ & $-5.1$
    & $-10$ & $-4.9$
    & $-10$ & $-4.9$
    & $-11$ & $-5.3$
     \\
    & \multicolumn{1}{ c }{200}
    & $-10$ & $-4.0$
    & $-8.8$ & $-3.2$
    & $-7.8$ & $-2.7$
    & $-8.4$ & $-2.9$
    & $-8.0$ & $-3.0$
     \\
    & \multicolumn{1}{ c }{300}
    & $-12$ & $-4.9$
    & $-11$ & $-4.0$
    & $-11$ & $-3.9$
    & $-9.9$ & $-3.8$
    & $-11$ & $-3.9$
     \\
    % \clineB{1-12}{2}
    \midrule
    {\multirow{4}{*}{$\nbasis = 4 \nlecs$} } 
    & \multicolumn{1}{ c }{5}
    & $-10$ & $-7.3$
    & $-8.8$ & $-6.4$
    & $-8.8$ & $-6.1$
    & $-8.5$ & $-6.4$
    & $-8.3$ & $-6.1$
     \\
    & \multicolumn{1}{ c }{100}
    & $-13$ & $-6.5$
    & $-12$ & $-5.3$
    & $-11$ & $-5.1$
    & $-11$ & $-5.0$
    & $-11$ & $-5.4$
     \\
    & \multicolumn{1}{ c }{200}
    & $-10$ & $-4.4$
    & $-9.3$ & $-3.6$
    & $-8.5$ & $-3.0$
    & $-8.8$ & $-3.3$
    & $-8.8$ & $-3.3$
     \\
    & \multicolumn{1}{ c }{300}
    & $-12$ & $-5.1$
    & $-11$ & $-4.0$
    & $-10$ & $-4.1$
    & $-10$ & $-3.8$
    & $-11$ & $-4.0$
\end{tabular}
\end{ruledtabular}
\end{table*}

The speed-up between the emulators and the simulator is highly implementation dependent (e.g., to-be-considered factors are the desired accuracy, idiosyncrasies of the solver, programming language, level of parallelization, hardware, etc.). The emulator speed-up will depend on the size of the quadrature mesh used by the simulator to obtain the high-fidelity solution. For reproducing the total cross section using the NVP emulator, Ref.~\cite{Melendez:2021lyq} states an emulator speed-up factor of $>300\times$ faster than the simulator in CPU time. When doubling the quadrature mesh size this factor becomes $>1000\times$. When comparing the KVP and NVP emulator speeds using one boundary condition (no anomaly checking) for the ${^1}S_0$ uncoupled partial wave, the KVP emulator is slightly slower due to the Lagrange multiplier in Eq.~\eqref{eq:coeff_solution} and numerical operations needed to solve Eq.\eqref{eq:kvp_trial_functional}. Mitigation of Kohn anomalies (by emulating multiple boundary conditions) will further contribute to slowing down the KVP emulator, or any other emulator.

%%%%%%%%%%%%%%%%%%%%%%%%%%%%%%%%%%%%%%%%%%%%%%%%%%%%%%%%
\section{Summary and outlook}
\label{sec:summary}
%%%%%%%%%%%%%%%%%%%%%%%%%%%%%%%%%%%%%%%%%%%%%%%%%%%%%%%%

We showed that the coordinate space KVP emulator for $NN$ scattering~\cite{Furnstahl:2020abp,Drischler:2021qoy} can be extended to momentum space and coupled channels, and demonstrated its efficiency in accurately reproducing phase shifts and \npr\ observables using a modern chiral interaction at N$^4$LO$+$. In addition, we provided two methods to implement the emulator, with the Gl{\"o}ckle spline interpolation method having a faster offline stage, but less accurate online stage than the standard method. By emulating (independent) scattering solutions associated with different asymptotic boundary conditions in each partial wave and weighting the results (e.g., for the $S$ matrix), spurious singularities known as Kohn anomalies were successfully mitigated for the KVP-based emulators~\cite{Drischler:2021qoy}.

We also constructed an NVP-based emulator and assessed how well the three emulators reproduced their respective high-fidelity solution for the ${^1}S_0$ and ${^3}S_1$--${^3}D_1$ partial waves, total and differential cross sections, and analyzing powers. While all emulators produced errors well below experimental errors~\cite{Perez:2013mwa}, the KVP standard method and NVP emulators most closely reproduced the simulator, while the KVP Gl{\"o}ckle spline interpolation emulator was overall the least accurate. The KVP emulator was found to have a slower online stage than the NVP emulator because it has to evaluate a higher-dimensional matrix and perform overall more numerical operations. We stress, however, that the emulators' speed-ups are highly implementation dependent and should be further investigated. Extensions of the NVP-based emulator for anomaly mitigation with minimal computational cost, similar to the KVP-based emulators, should also be investigated~\cite{Melendez:2021lyq}. An alternative procedure for mitigating anomalies would be constructing the estimated $S$ matrix using solutions from emulators based on different variational principles, as opposed to emulating multiple boundary conditions. Reference~\cite{Drischler:2022ipa} provides further perspectives regarding different emulators (KVP- and NVP-based included) and efficient offline-online decompositions.

Although we considered here only \chiEFT\ $NN$ potentials for \npr\ scattering, the constructed emulators are generally applicable to two-body scattering, including $pp$ scattering and nuclear reactions with complex-valued optical potentials. 
To help implement these fast \& accurate scattering emulators in Bayesian parameter estimations, we provide self-contained set of codes that generate all results and figures shown in this paper~\cite{BUQEYEsoftware}. 
Furthermore, we have written a pedagogical introduction to projection-based emulators~\cite{Drischler:2022ipa} with interactive, open-source \texttt{PYTHON} code to facilitate implementations of fast \& accurate emulators even further. 
However, taking full advantage of emulators for uncertainty quantification in nuclear scattering and reaction calculations will require generalizations to higher-body scattering and non-affine potentials. Recent advances in this direction are already promising~\cite{Zhang:2021jmi}.

%%%%%%%%%%%%%%%%%%%%%%%%%%%%%%%%%%%%%%%%%%%%%%%%%%%%%%%%
\begin{acknowledgments}
We thank Evgeny Epelbaum for sharing a code that generates the SMS chiral potentials, Kyle Wendt for sharing a code that generates the spin observables, and Filomena Nunes for fruitful discussions. 
This work was supported in part by the National Science Foundation Award Nos. PHY-1913069 and PHY-2209442 and the NSF CSSI program under award
no. OAC-2004601 (BAND Collaboration~\cite{BAND_Framework}), and the NUCLEI SciDAC Collaboration under U.S. Department of Energy MSU subcontract no. RC107839-OSU\@.
This material is based upon work supported by the U.S. Department of Energy, Office of Science, Office of Nuclear Physics, under the FRIB Theory Alliance Award No. DE-SC0013617. 
\end{acknowledgments}

%%%%%%%%%%%%%%%%%%%%%%%%%%%%%%%%%%%%%%%%%%%%%%%%%%%%%%%%
\appendix

\section{Mitigating Kohn anomalies} \label{sec:anomalies}
We follow the method developed in Ref.~\cite{Drischler:2021qoy} to detect and mitigate Kohn anomalies (see also Ref.~\cite{Lucchese:1989zz}). The estimated $S$ matrix is calculated from the emulator solutions by using a weighted sum of averages. Letting $L_1$ and $L_2$ be two independent KVP functional solutions, this weighted sum is computed by first calculating the relative residuals
\begin{align} \label{eq:gamma_condition}
  \gamma_{\mathrm{rel}}(L_1, L_2) = \mathrm{max} 
        \Bigg\{ 
               \Bigg| \frac{S(L_1)}{S(L_2)} - 1 \Bigg|, 
               \Bigg| \frac{S(L_2)}{S(L_1)} - 1 \Bigg|
        \Bigg\},
\end{align}
for all emulated KVP solutions without repetitions to avoid the trivial case where $L_1 = L_2$. Using a consistency check, $\gamma_{\mathrm{rel}} < \epsilon_{\mathrm{rel}}$, with $\epsilon_{\mathrm{rel}} = 10^{-1}$, we select the set of pairs $\mathcal{P} = \{(L_1, L_2)\}$ that satisfies this check. If at least one consistency check passes, the $S$ matrix is now estimated by the weighted sum of averages
\begin{align} \label{eq:mixed_S}
  [S]^{\mathrm{(mixed)}}_{\mathrm{KVP}} 
    &= \sum_{\mathrm{(L_1, L_2)} \in \mathcal{P}} \omega(L_1, L_2) \frac{S(L_1) + S(L_2)}{2}, \\
  \omega(L_1, L_2) 
    &= \frac{\gamma_{\mathrm{rel}}(L_1, L_2)^{-1}}{\sum_{\mathrm{(L'_1, L'_2)} \in \mathcal{P}} \gamma_{\mathrm{rel}}(L'_1, L'_2)^{-1}}.
\end{align}
If no consistency check passes, one could change the basis size to shift the position of the Kohn anomalies in the parameter space. However, we found that using Eq.~\eqref{eq:mixed_S} was sufficient to mitigate Kohn anomalies in our applications.

We first calculate Eq.~\eqref{eq:delta_U_general} using Eq.~\eqref{eq:momentum_space_wf_coupled}, then rescale Eq.~\eqref{eq:delta_U_general} using the relations from Appendix~B of Ref.~\cite{Drischler:2021qoy},
\begin{align}
    \dU^{ (\umatrix')}
    &=
    C^{'-1}(L_i) \, C^{'-1} (L_j) \frac{\mathrm{det} \, \umatrix}{\mathrm{det} \, \umatrix'} \dU^{ (\umatrix)}, \\
    C'(L) 
    &= \frac{\mathrm{det} \, \umatrix}{\mathrm{det} \, \umatrix'} \frac{u'_{11} - u'_{10} K(L)}{u_{11} - u_{10} K(L)}.
\end{align}
Here, $\umatrix$ and $\umatrix'$ are nonsingular matrices parametrizing the scattering boundary conditions; the $K$, $K^{-1}$, and $T$ scattering matrices, respectively, are given by
\begin{align}
    \umatrix_{K} = 
    \begin{pmatrix}
        1 & 0 \\
        0 & 1
    \end{pmatrix},
    \quad
    \umatrix_{K^{-1}} = 
    \begin{pmatrix}
        0 & 1 \\
        1 & 0
    \end{pmatrix},
    \quad
    \umatrix_T = 
    \begin{pmatrix}
        1 & 0 \\
        i & 1
    \end{pmatrix}.
\end{align}
The $\umatrix$ matrix parametrizes the initial boundary condition associated with $L$, while the $\umatrix'$ parametrizes the final boundary condition associated with $L'$.

The snapshots used in the emulator's offline stage are transformed using the M{\"o}bius transform~\cite{Drischler:2021qoy}
\begin{align} \label{eq:moebius_L}
    L'(L) =  \frac{-u'_{01}+u'_{00} K(L)}{u'_{11} -u'_{10}K(L)} .
\end{align}
Once we obtain an emulator solution, we transform that solution back into its $K$ matrix form using
\begin{align}
    K(L) = \frac{u_{01} + u_{11} L}{u_{00} + u_{10} L} .
\end{align}
For the estimated $S$ calculation, the KVP solution pairs $(L_1, L_2)$ being evaluated are the $K$ matrix solutions obtained from the different boundary conditions used [e.g., $\gamma_{\mathrm{rel}}(K(K), K(K^{-1}))$, $\gamma_{\mathrm{rel}}(K(K), K(T)$), and $\gamma_{\mathrm{rel}}(K(K^{-1}), K(T)$)]. See Ref.~\cite{Drischler:2021qoy} for more details.

\section{Formalism details}
\label{sec:coupled_channel_details}

Here we provide clarifying remarks about how Eq.~\eqref{eq:kvp_trial_functional} arises in the coupled case.
In particular, we focus on two questions about the specific manner in which the coefficients $\vec{\weights}$ enter into Eq.~\eqref{eq:kvp_trial_functional}.
% Why can each $\mathcal{K}^{\ell'\ell}$ be emulated separately?
% Why does $\weights_i^{\ell'\ell}$ show up quadratically even when the trial function $\ket{\psi_j^{s\ell}}$ has an $\ell$ that differs from the $\ell'$ in $\bra{\psi_i^{\ell's'}}$?
% Should not each of $\ket{\psi^{s\ell}}$ and $\bra{\psi^{\ell's'}}$ have its own basis expansion with their own independent coefficients?
\\

\textit{Why can $\mathcal{L}^{s s'}$ be emulated separately for each $s s'$ pair rather than with one global set of coefficients for the coupled block?} 

%To answer this question, we begin to build intuition in the uncoupled case.
For uncoupled channels, each partial wave is independent of one another, thus they can be emulated individually using trial wave functions and coefficients that are specific to the channel under consideration.
Without loss of generality, let us consider two uncoupled channels labeled as $s = 0$ and $s = 1$, and let $\vec{\weights}^{(0)}$ and $\vec{\weights}^{(1)}$ denote the independent sets of coefficients found by making each channel's KVP stationary.
%When two channels are uncoupled, it is natural to emulate each separately, using trial wave functions and coefficients that are specific to the channel under consideration.
%When solving for a particular Hamiltonian, the coefficients $\{\weights_i^{(0)}\}$ could be quite different than that of $\{\weights_i^{(1)}\}$. 
To move toward the coupled regime, imagine adiabatically turning on the coupling between these two originally uncoupled channels.
The coefficients for each channel should remain nearly fixed to their previously uncoupled values, 
%where $\vec{\weights}^{(0)} \neq \vec{\weights}^{(1)}$ in general,
but the coupling will introduce a new set of coefficients $\vec{\weights}^{(01)} \neq \vec{\weights}^{(0)} \neq \vec{\weights}^{(1)}$ that must be determined.
Hence, each independent channel in the coupled case will have its own set of coefficients. 
Attempting to force a global set of coefficients for a coupled system would be inconsistent with the treatment in the uncoupled case and also degrade accuracy in general.
A more technical answer follows from the \mbox{(Petrov-)}Galerkin procedure described below.
\\

\textit{Should not each of $\ket{\psi^{s'}}$ and $\bra{\psi^{s}}$ have its own basis expansion with their own independent coefficients?}

No, there is only one set of coefficients that enter quadratically in Eq.~\eqref{eq:kvp_trial_functional}.
A way of understanding how the coefficients enter in Eq.~\eqref{eq:kvp_trial_functional} follows from the \mbox{(Petrov-)}Galerkin orthogonalization procedure (see also Ref.~\cite{Drischler:2022ipa}).
Rather than starting with a variational principle, the \mbox{(Petrov-)}Galerkin approach starts with the Schr\"odinger equation.
Like the variational approach, it expands $\ket{\psi^{s'}}$ as a linear combination of known functions, but determines the basis coefficients by enforcing orthogonality against a set of \emph{test functions}.
%To arrive at the KVP used in this work, we choose the test function to be the full wave function $\ket{\psi^{s'}}$, and obtain an emulator for $L^{s s'}_E$ by enforcing a particular boundary condition on its asymptotic form.
%The resulting set of equations is equivalent to those that follow from making the KVP stationary.
%Thus by following the Galerkin procedure we can infer how the coefficients are to enter in Eq.~\eqref{eq:kvp_trial_functional}.
For the diagonal channels, the test functions are chosen to have the same exit channel as the trial functions (standard Galerkin approach). On the other hand, the test functions for the off-diagonal channels are chosen to have a different exit channel ($s$) than the trial functions ($s'$) (Petrov-Galerkin approach).
%To arrive at the KVP used in this work, we choose the test function to be the full wave function $\ket{\psi^{s'}}$, and obtain an emulator for $L^{s s'}_E$ by enforcing a particular boundary condition on its asymptotic form.
The resulting set of linear equations is equivalent to those that follow from making the KVP stationary for each combination of $(s',s)$ independently.
Thus by following the \mbox{(Petrov-)}Galerkin procedure we can determine how the coefficients are to enter in Eq.~\eqref{eq:kvp_trial_functional}.

We now show how a Petrov-Galerkin procedure can be used to determine the KVP coefficients. This discussion will follow closely that of Ref.~\cite{Drischler:2022ipa}, however using coupled-channel notation and more general boundary conditions consistent with the general KVP.\@
Starting from (the strong form of) the Schr\"odinger equation
%
% \begin{align}
%     \braket{\chi^{st} | [H - E]^{tt'} | \chi^{t's'}} = - \braket{\chi^{st} | V^{ts'} | \phi^{s'}} ,
% \end{align}
\begin{equation} 
    \widehat H(\param) \ket{\psi^{s'}} = E  \ket{\psi^{s'}},
\end{equation}
we can derive its weak form after multiplying by a test function $\bra{\psi^{s}}$
\begin{align} \label{eq:kohn-weak-initial}
    \braket{\psi^{s} | \widehat H(\param) - E | \psi^{s'}} = 0.
\end{align}
This can be considered a \emph{Petrov-Galerkin} approach because $s \neq s'$ in general.
The boundary conditions can be made explicit via the relationship
\begin{align} \label{eq:int-by-parts-wronskian}
    0
    & = \braket{\psi^{s} | \widehat H(\param) - E | \psi^{s'}} \notag\\
    & = \braket{\psi^{s} | \widehat H^\dagger(\param) - E | \psi^{s'}} - \sum_{t} \left.\frac{W(r\psi^{ts}, r\psi^{t s'}; r)}{2\mu} \right|_{r=0}^{\infty},
\end{align}
where $\widehat H^\dagger$ denotes the operator acting to the left (via integration by parts) and where we have used $\psi^{ts}(r) = \braket{rt | \psi^s} = \braket{\psi^s | rt}$ and defined the Wronskian
\begin{align} \label{eq:wronskian}
    W(\phi, \psi; r) \equiv \phi(r) \psi'(r) - \phi'(r)\psi(r).
\end{align}
The wave function $r \psi$ vanishes at the origin, so that only the limit as $r\to\infty$ contributes.
By adding Eqs.~\eqref{eq:int-by-parts-wronskian} and~\eqref{eq:kohn-weak-initial}, we have
\begin{align} \label{eq:kohn-weak-full}
    \braket{\psi^{s} | \widehat H(\param) - E | \psi^{s'}} & + \braket{\psi^{s} | \widehat H^\dagger(\param) - E | \psi^{s'}} \notag\\
    & = \sum_{t} \left.\frac{W(r\psi^{ts}, r\psi^{t s'}; r)}{2\mu} \right|_{r=0}^{\infty}.
\end{align}
This is the weak form for general $\ket{\psi^{s'}}$ and $\bra{\psi^s}$.
We can arrive at the discrete form by inserting basis states $\ket{\psi_i^s}$ that satisfy the asymptotic boundary conditions
\begin{align} \label{eq:free-sol-coordinate}
\psi^{st}(r) \xrightarrow[r\to\infty]{}  \delta_{st}\bar{\phi}_s^{\text{(0)}}(r) + L^{st} \, \bar{\phi}_s^{\text{(1)}}(r) \, ,
\end{align}
where 
\begin{equation}\label{eq:freeSolMatch}
\begin{pmatrix}
    \bar{\phi}_{\ell}^{\text{(0)}}(r)\\
    \bar{\phi}_{\ell}^{\text{(1)}}(r)
\end{pmatrix} \propto %=
%\mathcal{N}^{-1}
\begin{pmatrix}
    u_{00} & u_{01}\\
    u_{10} & u_{11}\\
\end{pmatrix}
\begin{pmatrix}
     j_\ell(qr) \\
    \eta_\ell(qr)
\end{pmatrix} \,.
\end{equation}
With this substitution, we have, for $i \in [1, \nbasis]$,
\begin{align}
    \dU^{ss'}_{ij} \weights_j =  L^{ss'}_i \sum_j\weights_j - L^{s's}_j \weights_j,
\end{align}
where the expression for $\dU^{ss'}_{ij}$ is given by Eq.~\eqref{eq:delta_U_general}.
We must now implement the constraint $\sum_j\weights_j = 1$, which is performed here by a Lagrange multiplier $\lambda$ mimicking a variational approach (see Ref.~\cite{Melendez:2022kid} for details):
\begin{align} \label{eq:kohn-weak-reduced-unfinished}
    \lambda + \dU^{ss'}_{ij} \weights_j =  L^{ss'}_i\sum_j\weights_j - L^{s's}_j \weights_j.
\end{align}
The sum multiplying $L^{ss'}_i$ can be evaluated using the constraint $\sum_j \weights_j=1$, and we can make the redefinition $\lambda' \equiv \lambda + \sum_j \weights_j L_j^{s's}$ without impacting the solution because this term does not depend on $i$.
Thus, we have
\begin{align} \label{eq:kohn-weak-reduced}
    \lambda' - \vec{L}(E) + \dU \kvpweights = 0,
\end{align}
which is exactly Eq.~\eqref{eq:coeff_solution} found by making the KVP stationary.
This simplification can be understood by noting that if $\{\kvpweights, \lambda_\star\}$ satisfy Eq.~\eqref{eq:kohn-weak-reduced-unfinished}, then we know that $\{\kvpweights, \lambda_\star'\}$ is the unique solution to Eq.~\eqref{eq:kohn-weak-reduced}.
Therefore, we can solve Eq.~\eqref{eq:kohn-weak-reduced} to obtain $\kvpweights$ rather than Eq.~\eqref{eq:kohn-weak-reduced-unfinished}.
In conclusion, using the Petrov-Galerkin projection of the homogeneous Schrödinger equation with trial and test bases of $\ket{\psi_i^{s'}}$ and $\bra{\psi_i^{s}}$, respectively, we were able to obtain the same coefficients as the KVP in Eq.~\eqref{eq:coeff_solution}, which yield the same on-shell $L^{ss'}$ matrix when used in Eq.~\eqref{eq:kvp_trial_functional}.

\section{KVP emulator construction details}
\label{sec:emulator_details}
For single channel scattering over a $k \times p$ momentum grid using the $K$ matrix ($\det \umatrix = 1$), %Eq.~\eqref{eq:delta_U_general} becomes 
Eq.~\eqref{eq:delta_U_mom_representation} becomes
\begin{align} \label{eq:delta_U_mom_space_uncoupled}
    \dU_{ij}(\param)
    = \iint \limits_{0}^{\infty} \dd{k} \dd{p} 
    k^2 p^2 
    \bigl[
    \psi_i (k) V_{\param, j}(k,p) \psi_j (p) + (i \leftrightarrow j)
    \bigr],
\end{align}
with $V_{\param, j} (k,p)$ defined as in Eq~\eqref{eq:potential_definition}. We drop the superscripts for the uncoupled case since $s' = s$. Note that $\psi_i$ is \emph{not} complex conjugated. For the Gl\"{o}ckle method, one would simply substitute Eq.~\eqref{eq:momentum_space_wf_coupled} into Eq.~\eqref{eq:delta_U_mom_space_uncoupled} and interpolate the solutions to the integrals with the cubic spline polynomials $\mathcal{S}_k (\kzero)$. For the standard method, the Dirac $\delta$ distribution is analytically integrated; thus we obtain the following expression for $\dU_{ij}$
\begin{align} \label{eq:delta_U_integrals}
    \dU_{ij} (\param)
    = 
    V_{\param, j} (\kzero, \kzero) + \frac{2}{\pi} (I^1_{ij} + I^2_{ij}) + \frac{4}{\pi^2} I^3_{ij} + (i \leftrightarrow j),
\end{align}
with $I^1_{ij}$, $I^2_{ij}$, and $I^3_{ij}$ defined as
\begin{align}
    I^1_{ij} &= \mathbb{P} \int \limits_{0}^{\infty}\dd{k} \frac{k^2}{\kzero}
    \frac{K_i (\kzero, k)}{k^2 - k^2_0}
    V_{\param, j} (k, \kzero),\\
    I^2_{ij} &= \mathbb{P} \int \limits_{0}^{\infty}\dd{p} \frac{p^2}{\kzero}
    V_{\param, j} (\kzero, p) \frac{K_j (p, \kzero)}{p^2 - k^2_0},\\
    I^3_{ij} &= \mathbb{P} \iint \limits_{0}^{\infty}\dd{k}\dd{p} \frac{k^2 p^2}{\kzero^2} 
    \frac{K_i (\kzero, k)}{k^2 - k^2_0} V_{\param, j} (k, p) \frac{K_j (p, \kzero)}{p^2 - k^2_0}.
    \label{eq:kvp_integral_terms}
\end{align}

If $V$ has an affine dependence on the parameters $\param$, applying Eqs.~\eqref{eq:linear_params_pot} and \eqref{eq:delta_U_linear} produces
\begin{align}
    \dU^{0}_{ij} 
    &=  
    \iint \limits_{0}^{\infty} \dd{k} \dd{p} k^2 p^2 \bigl[
    \psi_i (k) V^0_{j} (k,p) \psi_j(p) + (i \leftrightarrow j) \bigr], \label{U0}
    \\    
    \Delta \boldsymbol{\widetilde U}^{1}_{ij} 
    &=
    \iint \limits_{0}^{\infty} \dd{k} \dd{p} k^2 p^2 \bigl[
    \psi_i (k) \boldsymbol{V}^1(k,p) \psi_j(p) + (i \leftrightarrow j) \bigl], \label{U1}
\end{align}
with
\begin{align}
    V^0_{j} (k,p) \equiv 2 \mu k_0 \big[
    V^0 (k,p) - V_j (k,p) \big].
\end{align}

For coupled-channel interactions ($s' \neq s$), the details of the emulation are more complex. In this case, we apply Eq.~\eqref{eq:kvp_trial_functional} to each individual channel in a partial wave, but the real difference lies in how Eq.~\eqref{eq:delta_U_general} is calculated. The usual way of solving for the phase shifts and mixing angle for the coupled channels involves building a $2 \times 2$ block matrix for the potential, 
\begin{align}
    V &= 
    \begin{pmatrix}
        V^{00} & V^{01} \\
        V^{10} & V^{11}
    \end{pmatrix}.
\end{align}
The same process can be applied to the emulator calculation when calculating Eq.~\eqref{eq:delta_U_general},
\begin{align}
    \dU = 
    \begin{pmatrix}
        \dU^{00} & \dU^{01} \\
        \dU^{10} & \dU^{11}
    \end{pmatrix}.
\end{align}
Each of the four blocks in $\dU$ has a separate functional, although there are contributions from the different wave functions and potentials (e.g., for the ${^3}S_1$--${^3}D_1$ partial wave $\dU^{ 00}$ depends on the ${^3}S_1$--${^3}S_1$, ${^3}S_1$--${^3}D_1$, and ${^3}D_1$--${^3}D_1$  wave functions and potentials).

Additionally, Eq.~\eqref{eq:momentum_space_wf_coupled} tells us that we can consider the momentum-space wave function for the individual channels $\psi^{st}$. Using Eq.~\eqref{eq:delta_U_mom_representation} with Eq.~\eqref{eq:potential_definition}, the functionals for the individual channels in a coupled-channel calculation (using the ${^3}S_1$--${^3}D_1$ as an example) will be
\begin{equation}
    \dU^{ss'}_{ij} = \iint^{\infty}_0 dk \, dp \, k^2 p^2 \bigl[\Delta u^{ss'}_{ij} + (i \leftrightarrow j) \bigr],
\end{equation}
with
\begin{align}
    \Delta u^{00}_{ij} 
    &= \psi^{00}_i 
    (V^{00}_{\param, j} \psi^{00}_j
    + V^{01}_{\param, j} \psi^{10}_j) \notag \\
    & \quad\null + 
    \psi^{10}_i
    (V^{10}_{\param, j} \psi^{00}_j
    + V^{11}_{\param, j} \psi^{10}_j),
    \label{eq:delta_U_mom_space_coupled_00} \\
    \Delta u^{01}_{ij} 
    &= \psi^{00}_i 
    (V^{00}_{\param, j} \psi^{01}_j
    + V^{01}_{\param, j} \psi^{11}_j) 
    \notag
    \\
    & \quad\null + 
    \psi^{10}_i
    (V^{10}_{\param, j} \psi^{01}_j
    + V^{11}_{\param, j} \psi^{11}_j),
    \label{eq:delta_U_mom_space_coupled_02} \\
    \Delta u^{10}_{ij} 
    &= \psi^{01}_i 
    (V^{00}_{\param, j} \psi^{00}_j
    + V^{01}_{\param, j} \psi^{10}_j) \notag
    \\ 
    & \quad\null +  
    \psi^{11}_i
    (V^{10}_{\param, j} \psi^{00}_j
    + V^{11}_{\param, j} \psi^{10}_j),
    \label{eq:delta_U_mom_space_coupled_20}
    \\
    \Delta u^{11}_{ij} 
    &= \psi^{01}_i 
    (V^{00}_{\param, j} \psi^{01}_j
    + V^{01}_{\param, j} \psi^{11}_j) 
    \notag \\
    & \quad\null + 
    \psi^{11}_i
    (V^{10}_{\param, j} \psi^{01}_j
    + V^{11}_{\param, j} \psi^{11}_j),
    \label{eq:delta_U_mom_space_coupled_22}
\end{align}
where we have suppressed the arguments for compactness. Note that the weights $\weights_i$ in Eq.~\eqref{eq:kvp_trial_functional} are different for each channel (i.e., $\dU^{ 00}$, $\dU^{11}$, and $\dU^{ 01} = \dU^{ 10}$), and are determined independently of one another. Once Eqs.~\eqref{eq:delta_U_mom_space_coupled_00} through~\eqref{eq:delta_U_mom_space_coupled_22} are calculated, the steps for the uncoupled channel calculation are applied to each $\dU^{ss'}_{ij}$ to obtain the emulator prediction, in particular Eqs.~\eqref{eq:delta_U_integrals} through~\eqref{eq:kvp_integral_terms}, and the separation of $\dU^{ss'} (\param)$ into parameter-dependent and parameter-independent pieces as described by Eq.~\eqref{eq:delta_U_linear}.

\section{Additional results} \label{sec:additional_resuls}

\begin{figure}[tb!]
    \centering
    \includegraphics
    {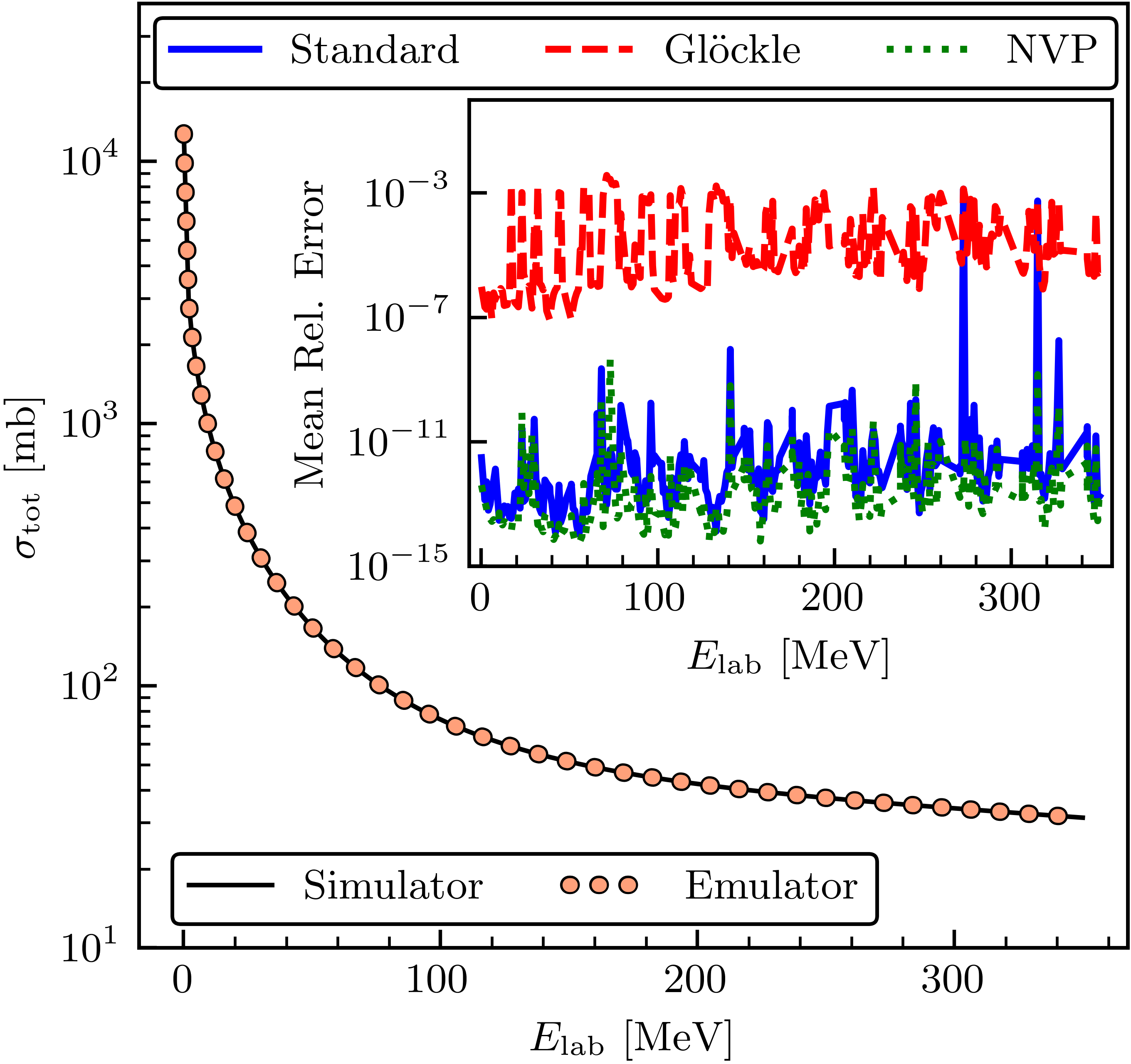}
    \caption{%
        As in Fig.~\ref{fig:cross_section}, but only emulating with the $K$ matrix. The mesh-induced spikes have been removed for this calculation.
     }%
    \label{fig:cross_section_spikes}%
\end{figure}

\begin{figure*}[tb]
    \centering
    \includegraphics
    {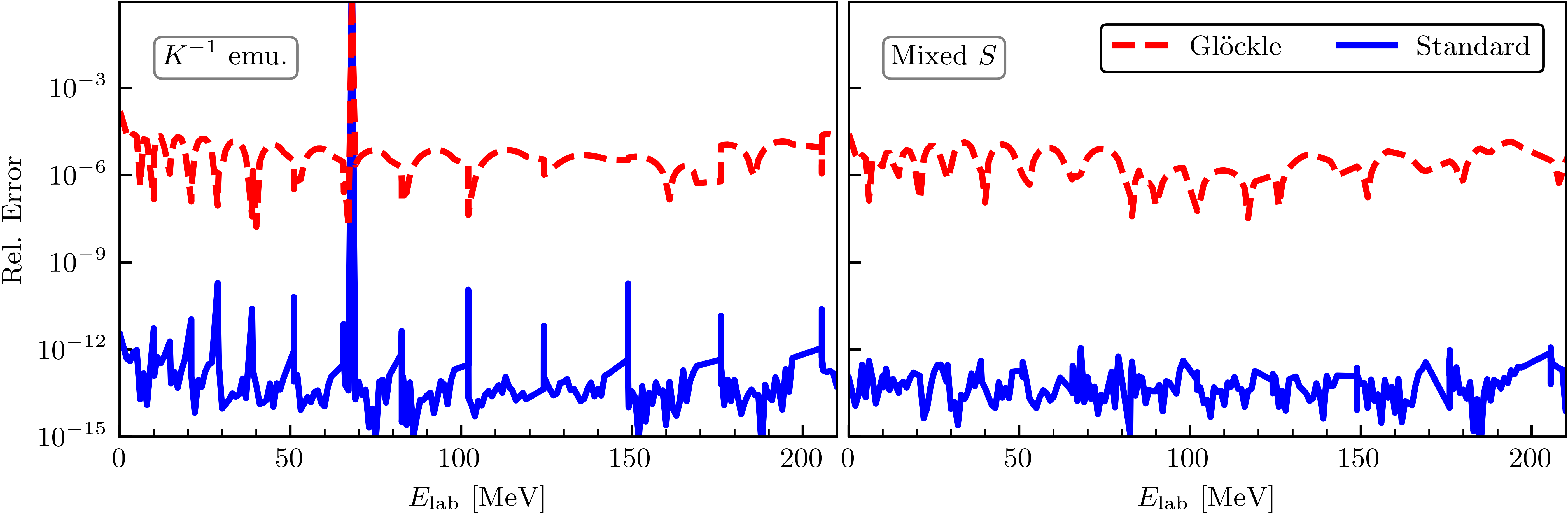}
    \caption{%
        Relative error of the ${^1}S_0$ channel for a basis size of $\nbasis = 2 \nlecs + 1$ for the N$^4$LO$+$ SMS potential with $\Lambda = 450 \MeV$ as a function of the laboratory energy.
        The left panel shows the relative error for an emulator using the $K^{-1}$ boundary condition. There is a Kohn anomaly at $E_\mathrm{lab}\approx 65$ MeV for both the standard and Gl{\"o}ckle emulators and mesh-induced spikes present throughout the energy grid.
        The right panel shows the relative error for the mixed $S$-matrix approach presented by Ref.~\cite{Drischler:2021qoy} with care taken to avoid the $\kzero$ values that correspond with a mesh point as described in Sec.~\ref{sec:results_partial_waves}.
        When comparing both graphs, the Kohn anomaly is no longer present and the mesh-induced spikes are much smaller in the right panel.
    }
    \label{fig:1S0_spikes_test}
\end{figure*}

Figure~\ref{fig:cross_section_spikes} shows the relative mean error for the total cross section using only the $K$ matrix boundary condition. Comparing to Fig.~\ref{fig:cross_section}, where we apply the weighted sum (mixed) $S$ approach, we see that for one boundary condition the relative mean error has Kohn anomalies (see $\Elab \approx 270 \MeV$ and $\approx 315 \MeV$ for the standard method and $\Elab \approx 40 \MeV$ and $\approx 130 \MeV$ for the Gl{\"o}ckle method) and a more spread-out error. From Fig.~\ref{fig:1S0_spikes_test} and comparing to Figs.~\ref{fig:cross_section} and~\ref{fig:cross_section_spikes}, we conclude that the mixed $S$ approach is indeed successful in mitigating the Kohn anomalies.

\begin{figure}[tb!]
    \centering
    \includegraphics
    {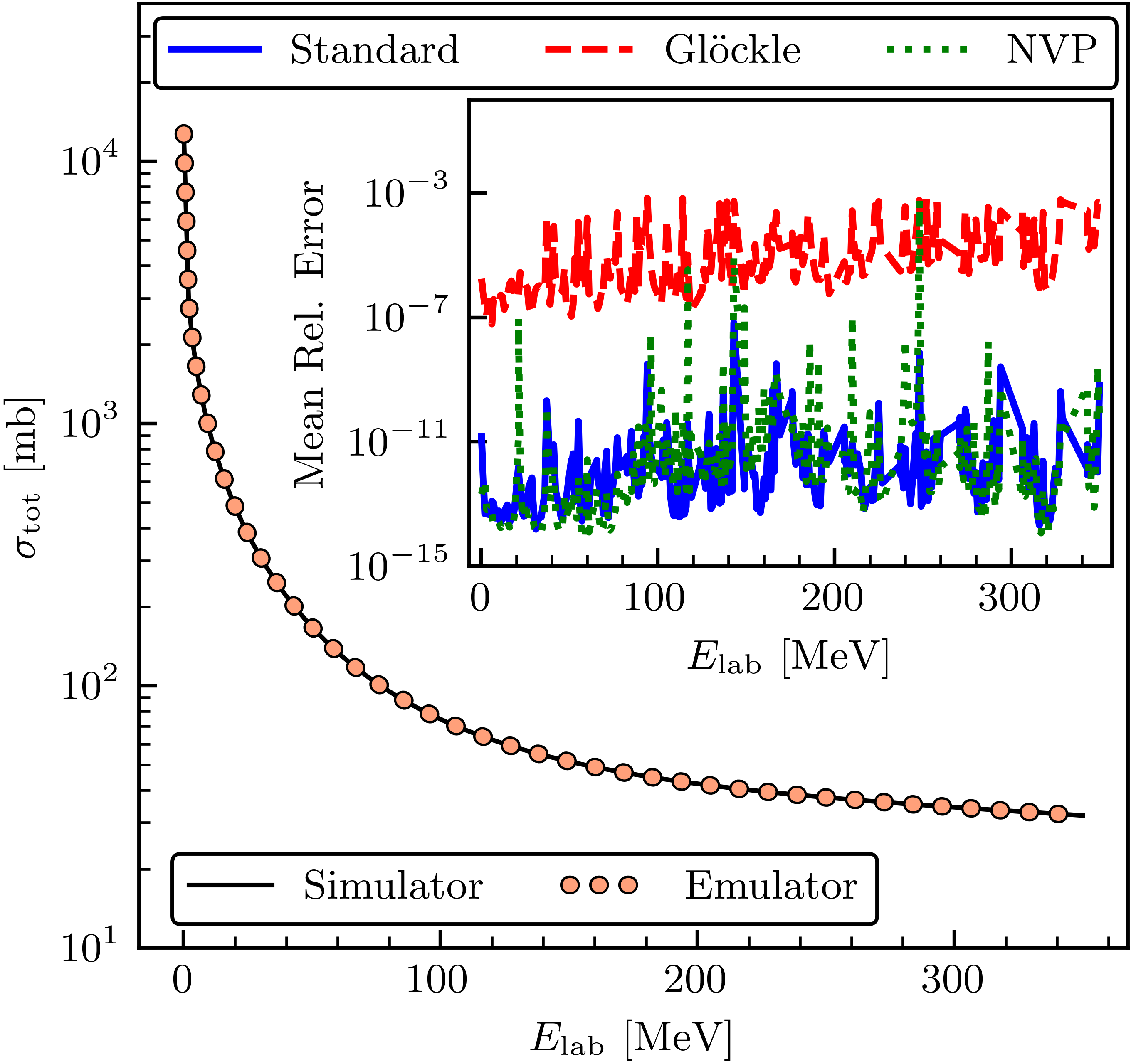}
    \caption{
        As in Fig.~\ref{fig:cross_section}, but for cutoff $\Lambda = 500 \MeV$.
     }%
    \label{fig:cross_section_500MeV}%
\end{figure}

Figure~\ref{fig:cross_section_500MeV} shows the relative mean error for the total cross section with momentum cutoff $500 \MeV$. The weighted sum (mixed) $S$ approach is used for the KVP emulator results. Here, the anomalies found in the NVP emulation are noticeable.

Figure~\ref{fig:1S0_spikes_test} shows the relative errors for the KVP emulators in the ${^1}S_0$ channel. The figure on the left shows the relative error when emulating with the $K^{-1}$ boundary condition and the one on the right shows the weighted sum (mixed) $S$ errors. In the figure on the left we can see a spike around $\Elab  \approx 65 \MeV$, which disappears when using the weighted sum $S$ approach. This is a clear example of the weighted sum $S$ approach helping to mitigate these anomalies. Additionally, there are other smaller mesh-induced spikes (i.e., not anomalies) present throughout the energy grid in the figure on the left that are \emph{not} in the figure on the right. These were mitigated by not allowing the $\kzero$ values to be close to any momentum mesh points. See Sec.~\ref{sec:results} for a more detailed description.

\begin{figure}[tb!]
    \centering
    \includegraphics
    {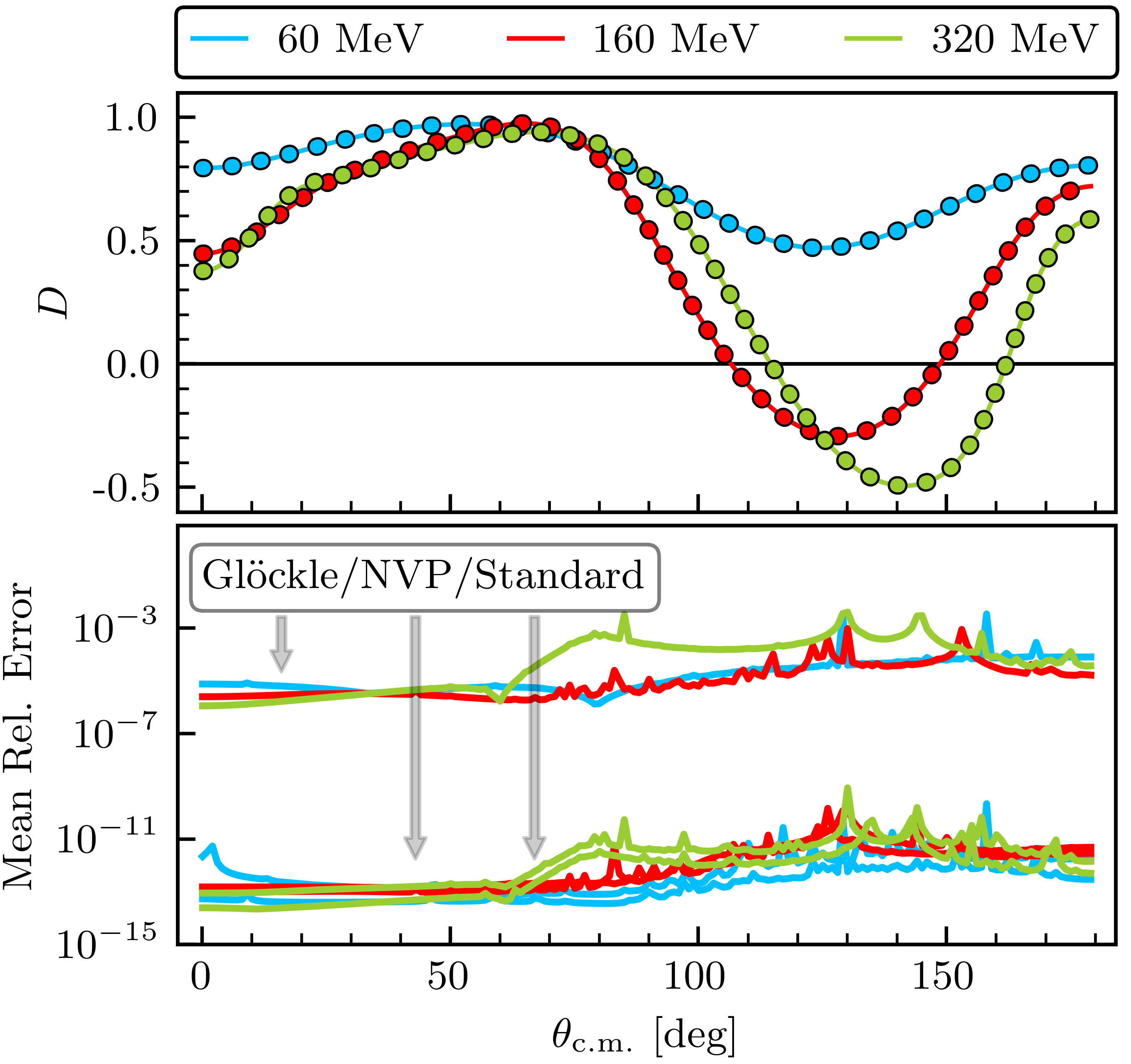}
    \caption{%
        As in Fig.~\ref{fig:differential_cs}, but for the depolarization $D$.
    }
    \label{fig:depolarization}
\end{figure}

\begin{figure}[tb!]
    \centering
    \includegraphics
    {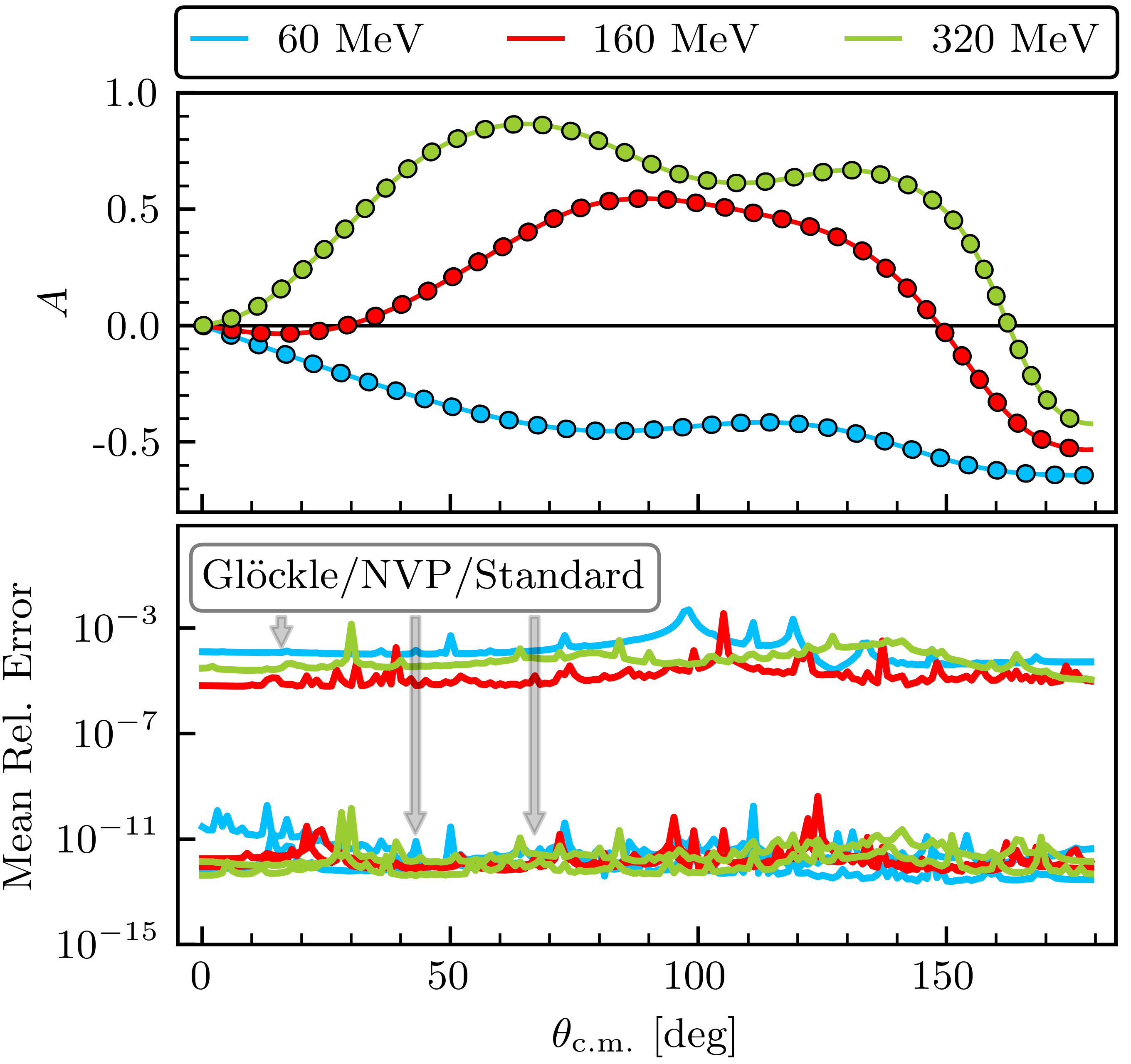}
    \caption{%
        As in Fig.~\ref{fig:differential_cs}, but for the spin-flip amplitude $A$.
    }
    \label{fig:spin_flip_amp_A}
\end{figure}

\begin{figure}[tb!]
    \centering
    \includegraphics
    {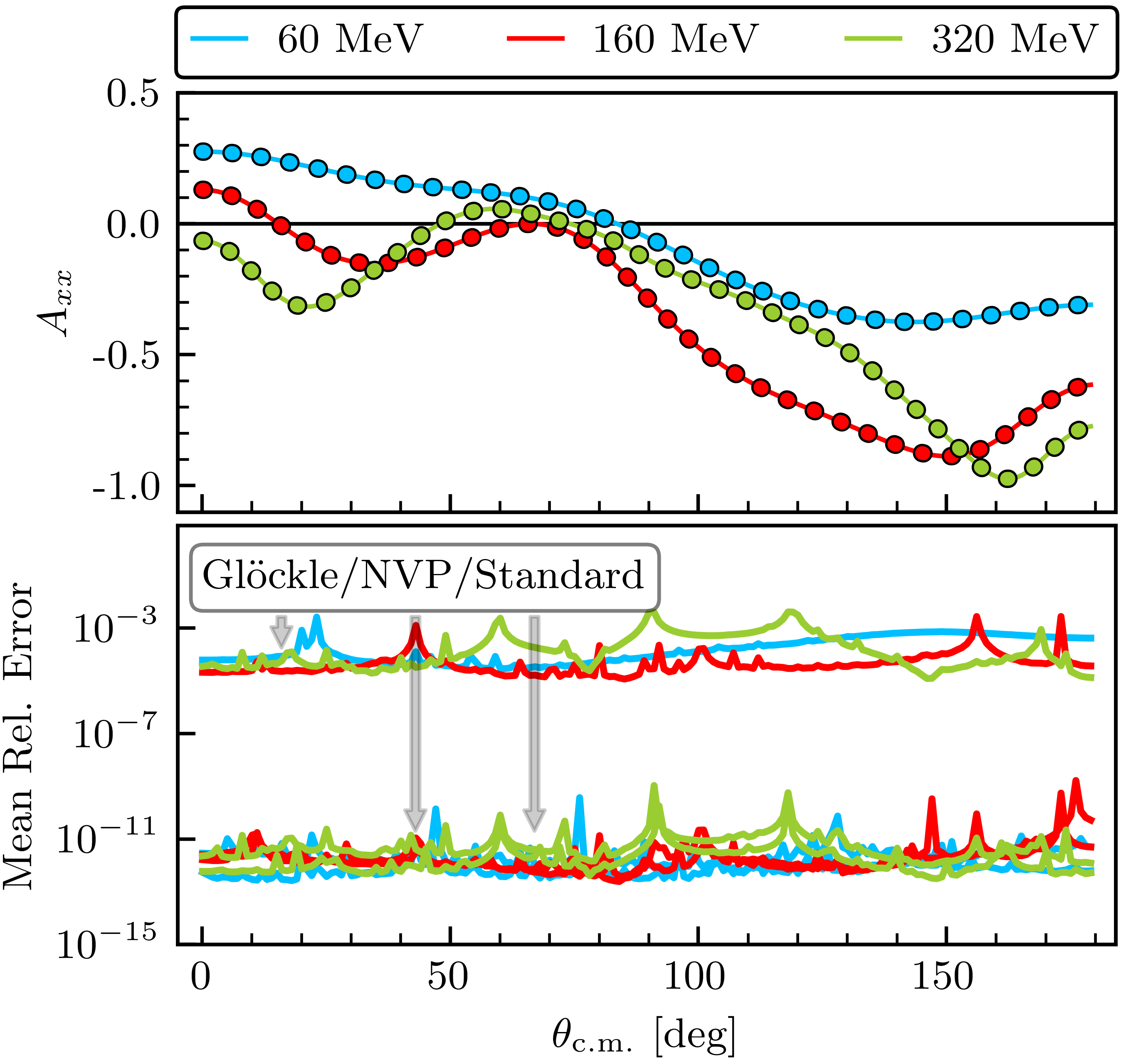}
    \caption{%
        As in Fig.~\ref{fig:differential_cs}, but for the spin-correlation amplitude $A_{xx}$.
    }
    \label{fig:spin_corr_amp_axx}
\end{figure}

\begin{figure}[tb!]
    \centering
    \includegraphics
    {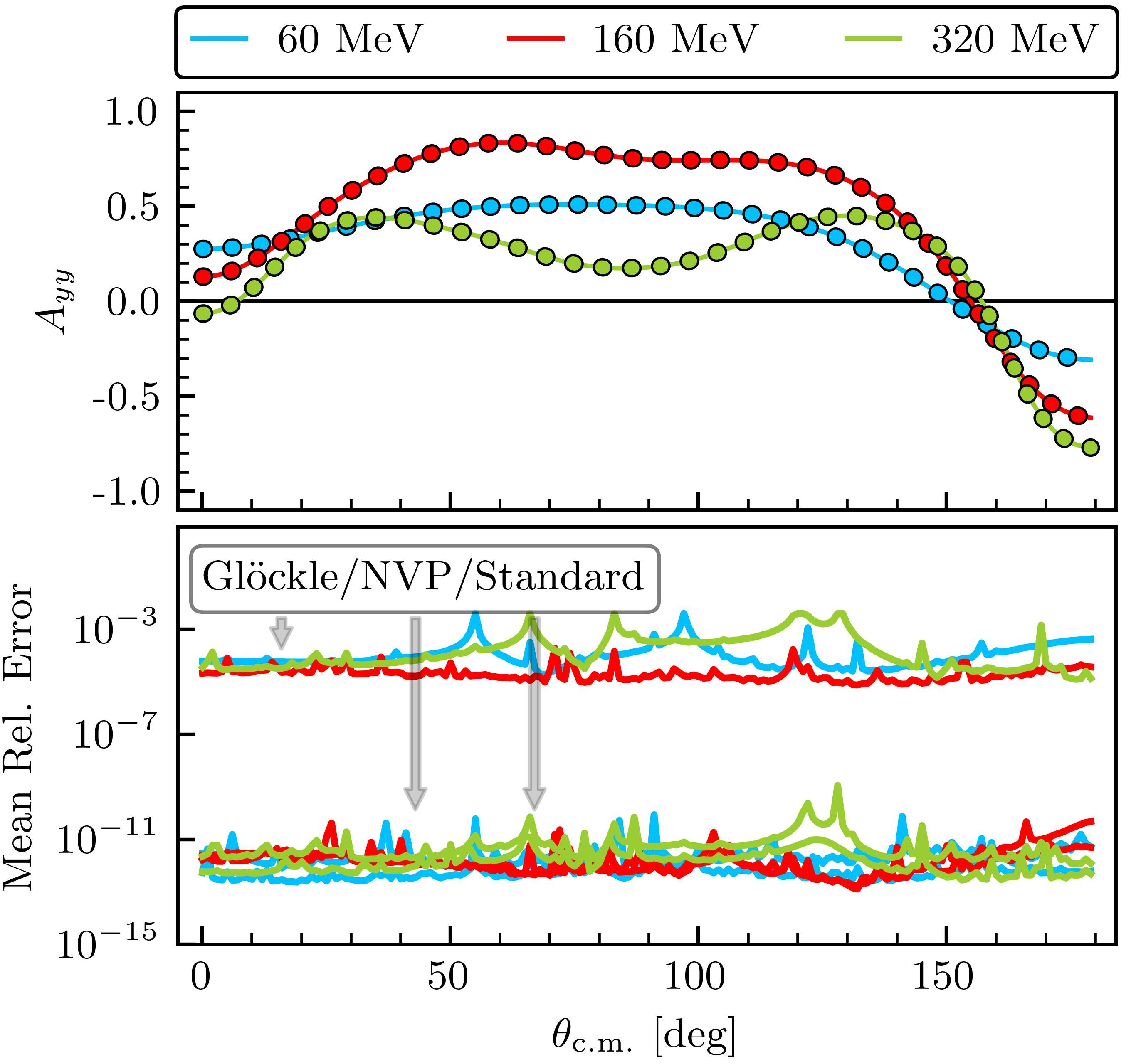}
    \caption{%
        As in Fig.~\ref{fig:differential_cs}, but for the spin-correlation amplitude $A_{yy}$.
    }
    \label{fig:spin_corr_amp_ayy}
\end{figure}

Figures~\ref{fig:depolarization} through~\ref{fig:spin_corr_amp_ayy} show emulator results for the following spin observables:
\begin{align} \label{eq:spin_obs_supp}
    \dv{\sigma}{\Omega} D = \, &\frac{1}{2} \Big[ |a|^2 + |b|^2 -|c|^2 -|d|^2 + |e|^2 + |f|^2 \Big] ,\\
    \dv{\sigma}{\Omega} A = \, &-\mathrm{Re} (a^* \, b - e^* \, f) \sin(\alpha + \frac{\theta}{2}) 
    \notag \\
    &\quad\null+ \mathrm{Re} (c^* \, d) \sin(\alpha - \frac{\theta}{2}) 
    \notag\\
    &\quad\null- \mathrm{Im} (b^* \, e + a^* \, f) \cos(\alpha + \frac{\theta}{2}) ,\\
    \dv{\sigma}{\Omega} A_{xx} = \, &\mathrm{Re} (a^* \, d) \cos(\theta) + \mathrm{Re} (b^* \, c) - \mathrm{Im} (d^* \, e ) \sin(\theta) , \\
    \dv{\sigma}{\Omega} A_{yy} = \, &\frac{1}{2} \Big[ |a|^2 + |b|^2 -|c|^2 -|d|^2 + |e|^2 + |f|^2 \Big],
\end{align}
where $D$ is the depolarization parameter, $A$ is the spin-flip amplitude, $A_{xx}$ and $A_{yy}$ are the spin-correlation amplitudes, and $\alpha$ a relativistic spin rotating angle that vanishes in the non-relativistic case~\cite{Melendez:2020xcs}. For identical particles, $f=0$. The results and conclusions are similar to those described in Sec.~\ref{sec:results_observables}.

\begin{figure}[tb!]
    \centering
    \includegraphics
    {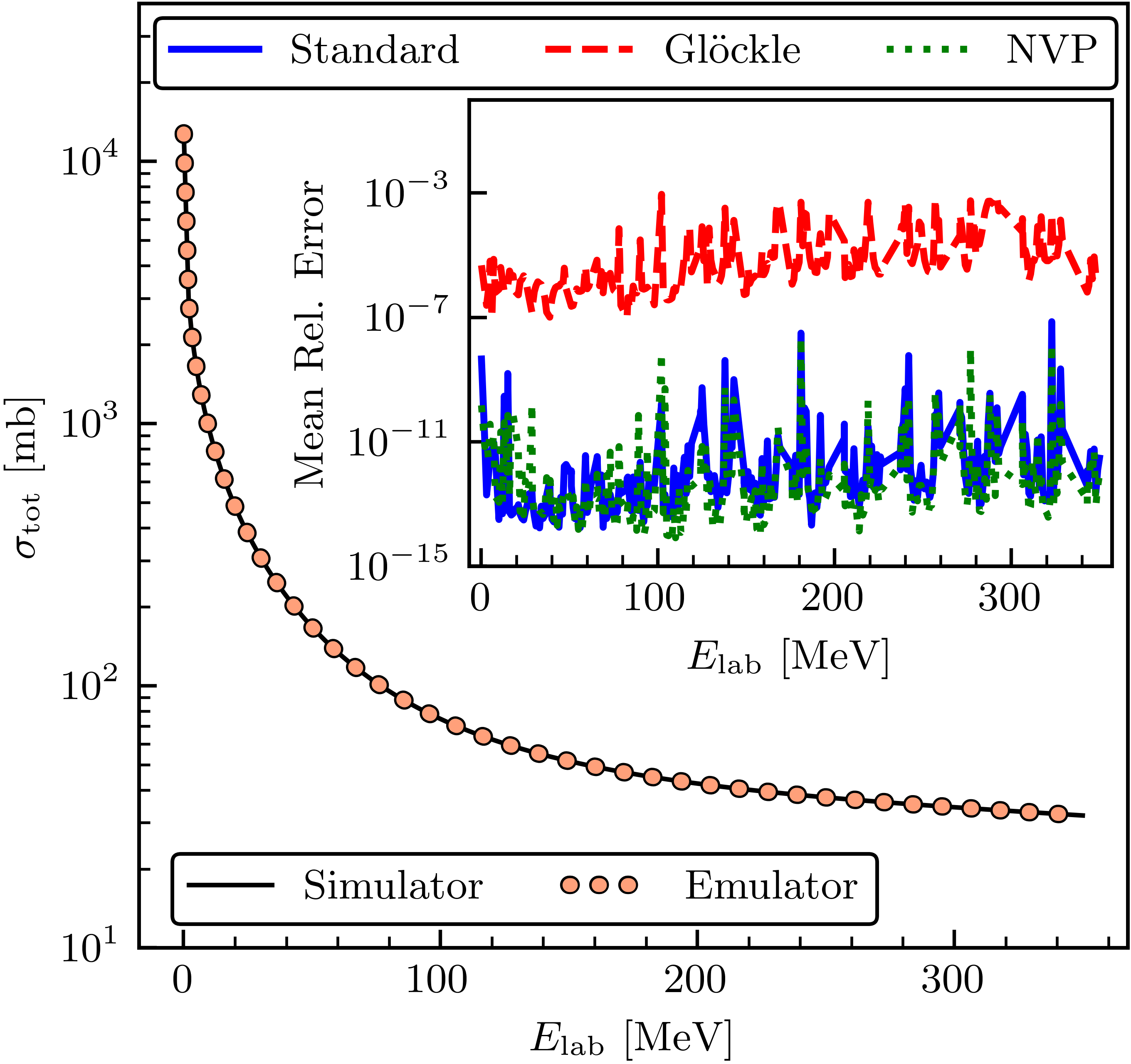}
    \caption{
        As in Fig.~\ref{fig:cross_section}, but for cutoff $\Lambda = 550 \MeV$.
     }%
    \label{fig:cross_section_550MeV}%
\end{figure}

Figure~\ref{fig:cross_section_550MeV} shows emulator results for the total cross section for the N$^4$LO$+$ SMS potential with momentum cutoff $550 \MeV$. The results and conclusions are similar to the ones described in the text for the $450 \MeV$ momentum cutoff (see Sec.~\ref{sec:results_observables}).

\begin{figure}[tb!]
    \centering
    \includegraphics
    {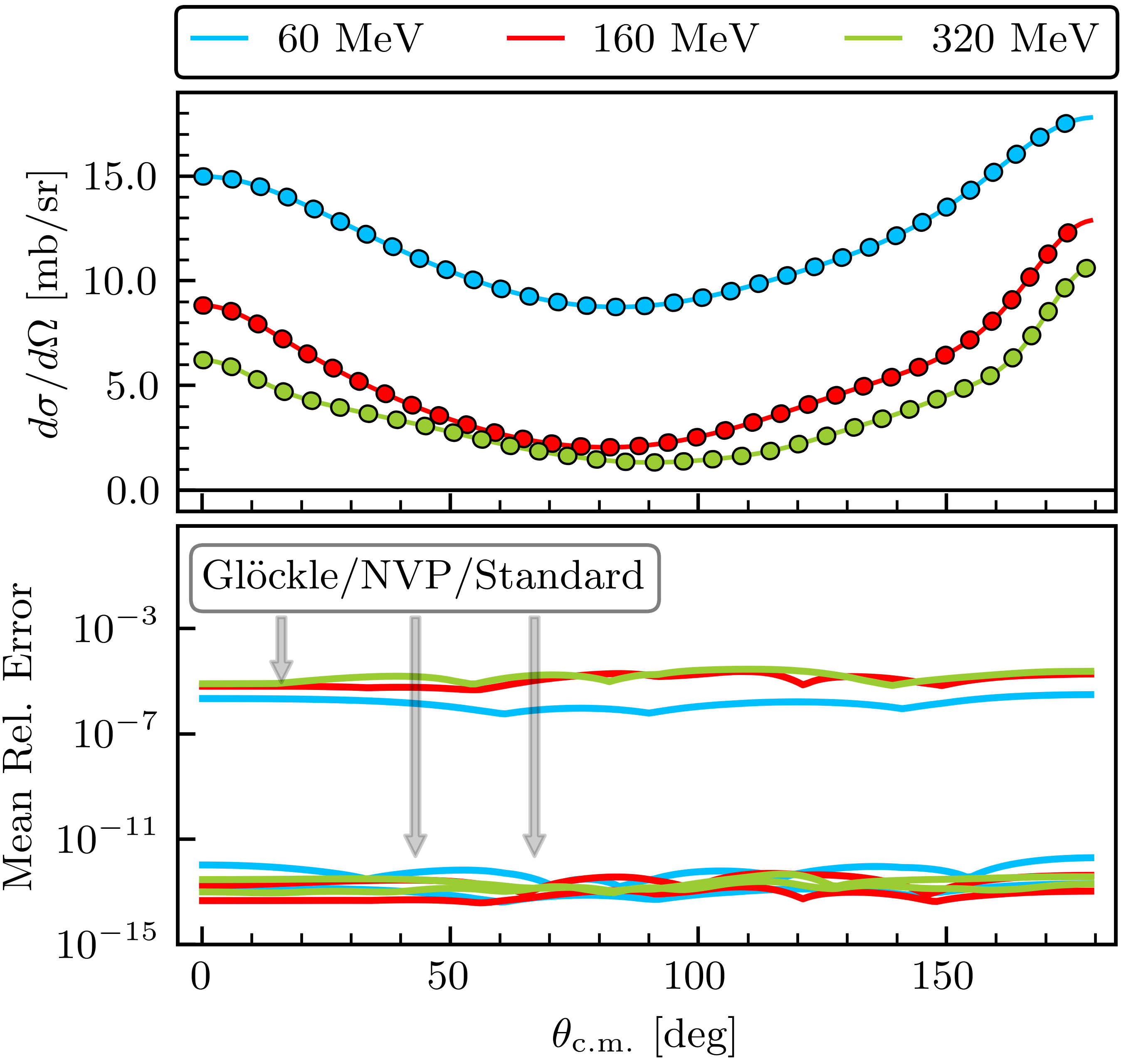}
    \caption{%
        As in Fig.~\ref{fig:differential_cs}, but for cutoff $\Lambda = 550 \MeV$.
    }
    \label{fig:differential_cs_550MeV}
\end{figure}

\begin{figure}[tb!]
    \centering
    \includegraphics
    {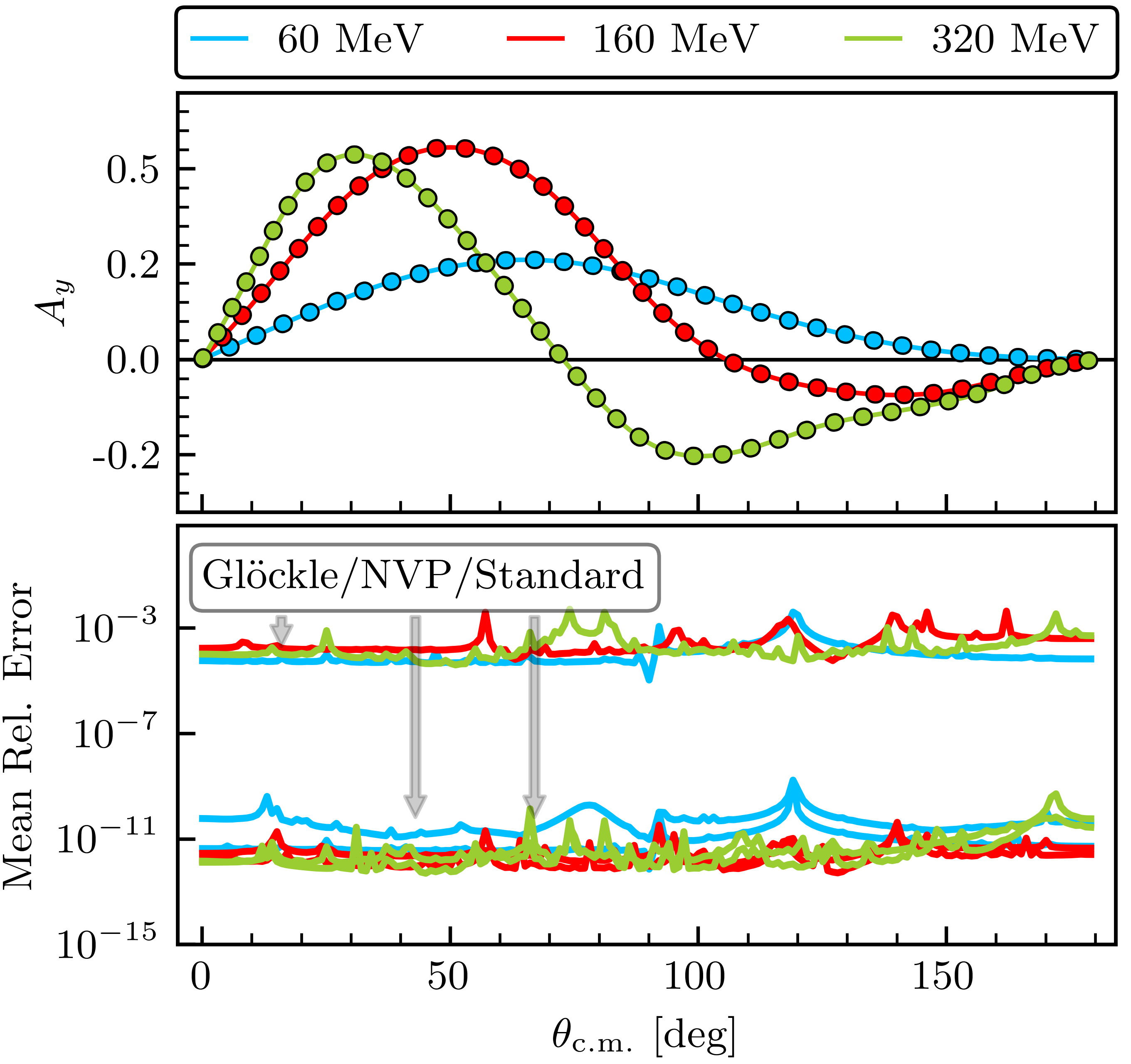}
    \caption{%
        As in Fig.~\ref{fig:spin_correlation_amp}, but for cutoff $\Lambda = 550 \MeV$.
    }
    \label{fig:spin_correlation_amp_550MeV}
\end{figure}

Figures~\ref{fig:differential_cs_550MeV} and~\ref{fig:spin_correlation_amp_550MeV} show emulator results for the differential cross section and analyzing power $A_y$ for the N$^4$LO$+$ SMS potential with momentum cutoff $550 \MeV$. The results and conclusions are similar to the ones described in the text for the $450 \MeV$ momentum cutoff (see Sec.~\ref{sec:results_observables}). These results and conclusions also extend down to momentum cutoff $400 \MeV$. The spin observables at $500 \MeV$ show larger errors on order of $10^{-7}$ for the NVP emulator at particular energies, which may come from Kohn anomalies at one or more of the sampled parameter sets (see Fig.~\ref{fig:cross_section_500MeV}); nevertheless, the errors are still well below experimental uncertainties~\cite{Perez:2013mwa}.

\clearpage
\bibliography{bayesian_refs}

\begin{thebibliography}{47}%
\makeatletter
\providecommand \@ifxundefined [1]{%
 \@ifx{#1\undefined}
}%
\providecommand \@ifnum [1]{%
 \ifnum #1\expandafter \@firstoftwo
 \else \expandafter \@secondoftwo
 \fi
}%
\providecommand \@ifx [1]{%
 \ifx #1\expandafter \@firstoftwo
 \else \expandafter \@secondoftwo
 \fi
}%
\providecommand \natexlab [1]{#1}%
\providecommand \enquote  [1]{``#1''}%
\providecommand \bibnamefont  [1]{#1}%
\providecommand \bibfnamefont [1]{#1}%
\providecommand \citenamefont [1]{#1}%
\providecommand \href@noop [0]{\@secondoftwo}%
\providecommand \href [0]{\begingroup \@sanitize@url \@href}%
\providecommand \@href[1]{\@@startlink{#1}\@@href}%
\providecommand \@@href[1]{\endgroup#1\@@endlink}%
\providecommand \@sanitize@url [0]{\catcode `\\12\catcode `\$12\catcode
  `\&12\catcode `\#12\catcode `\^12\catcode `\_12\catcode `\%12\relax}%
\providecommand \@@startlink[1]{}%
\providecommand \@@endlink[0]{}%
\providecommand \url  [0]{\begingroup\@sanitize@url \@url }%
\providecommand \@url [1]{\endgroup\@href {#1}{\urlprefix }}%
\providecommand \urlprefix  [0]{URL }%
\providecommand \Eprint [0]{\href }%
\providecommand \doibase [0]{https://doi.org/}%
\providecommand \selectlanguage [0]{\@gobble}%
\providecommand \bibinfo  [0]{\@secondoftwo}%
\providecommand \bibfield  [0]{\@secondoftwo}%
\providecommand \translation [1]{[#1]}%
\providecommand \BibitemOpen [0]{}%
\providecommand \bibitemStop [0]{}%
\providecommand \bibitemNoStop [0]{.\EOS\space}%
\providecommand \EOS [0]{\spacefactor3000\relax}%
\providecommand \BibitemShut  [1]{\csname bibitem#1\endcsname}%
\let\auto@bib@innerbib\@empty
%</preamble>
\bibitem [{\citenamefont {Epelbaum}\ \emph {et~al.}(2009)\citenamefont
  {Epelbaum}, \citenamefont {Hammer},\ and\ \citenamefont
  {Mei{\ss}ner}}]{Epelbaum:2008ga}%
  \BibitemOpen
  \bibfield  {author} {\bibinfo {author} {\bibfnamefont {E.}~\bibnamefont
  {Epelbaum}}, \bibinfo {author} {\bibfnamefont {H.-W.}\ \bibnamefont
  {Hammer}},\ and\ \bibinfo {author} {\bibfnamefont {U.-G.}\ \bibnamefont
  {Mei{\ss}ner}},\ }\bibfield  {title} {\bibinfo {title} {{Modern Theory of
  Nuclear Forces}},\ }\href {https://doi.org/10.1103/RevModPhys.81.1773}
  {\bibfield  {journal} {\bibinfo  {journal} {Rev. Mod. Phys.}\ }\textbf
  {\bibinfo {volume} {81}},\ \bibinfo {pages} {1773} (\bibinfo {year}
  {2009})},\ \Eprint {https://arxiv.org/abs/0811.1338} {arXiv:0811.1338}
  \BibitemShut {NoStop}%
%%CITATION = ARXIV:0811.1338;%%
\bibitem [{\citenamefont {Machleidt}\ and\ \citenamefont
  {Entem}(2011)}]{Machleidt:2011zz}%
  \BibitemOpen
  \bibfield  {author} {\bibinfo {author} {\bibfnamefont {R.}~\bibnamefont
  {Machleidt}}\ and\ \bibinfo {author} {\bibfnamefont {D.~R.}\ \bibnamefont
  {Entem}},\ }\bibfield  {title} {\bibinfo {title} {{Chiral effective field
  theory and nuclear forces}},\ }\href
  {https://doi.org/10.1016/j.physrep.2011.02.001} {\bibfield  {journal}
  {\bibinfo  {journal} {Phys. Rept.}\ }\textbf {\bibinfo {volume} {503}},\
  \bibinfo {pages} {1} (\bibinfo {year} {2011})},\ \Eprint
  {https://arxiv.org/abs/1105.2919} {arXiv:1105.2919} \BibitemShut {NoStop}%
%%CITATION = ARXIV:1105.2919;%%
\bibitem [{\citenamefont {Hammer}\ \emph {et~al.}(2020)\citenamefont {Hammer},
  \citenamefont {K\"onig},\ and\ \citenamefont {van Kolck}}]{Hammer:2019poc}%
  \BibitemOpen
  \bibfield  {author} {\bibinfo {author} {\bibfnamefont {H.-W.}\ \bibnamefont
  {Hammer}}, \bibinfo {author} {\bibfnamefont {S.}~\bibnamefont {K\"onig}},\
  and\ \bibinfo {author} {\bibfnamefont {U.}~\bibnamefont {van Kolck}},\
  }\bibfield  {title} {\bibinfo {title} {{Nuclear effective field theory:
  status and perspectives}},\ }\href
  {https://doi.org/10.1103/RevModPhys.92.025004} {\bibfield  {journal}
  {\bibinfo  {journal} {Rev. Mod. Phys.}\ }\textbf {\bibinfo {volume} {92}},\
  \bibinfo {pages} {025004} (\bibinfo {year} {2020})},\ \Eprint
  {https://arxiv.org/abs/1906.12122} {arXiv:1906.12122} \BibitemShut {NoStop}%
\bibitem [{\citenamefont {Epelbaum}\ \emph {et~al.}(2020)\citenamefont
  {Epelbaum}, \citenamefont {Krebs},\ and\ \citenamefont
  {Reinert}}]{Epelbaum:2019kcf}%
  \BibitemOpen
  \bibfield  {author} {\bibinfo {author} {\bibfnamefont {E.}~\bibnamefont
  {Epelbaum}}, \bibinfo {author} {\bibfnamefont {H.}~\bibnamefont {Krebs}},\
  and\ \bibinfo {author} {\bibfnamefont {P.}~\bibnamefont {Reinert}},\
  }\bibfield  {title} {\bibinfo {title} {{High-precision nuclear forces from
  chiral EFT: State-of-the-art, challenges and outlook}},\ }\href
  {https://doi.org/10.3389/fphy.2020.00098} {\bibfield  {journal} {\bibinfo
  {journal} {Front. Phys.}\ }\textbf {\bibinfo {volume} {8}},\ \bibinfo {pages}
  {98} (\bibinfo {year} {2020})},\ \Eprint {https://arxiv.org/abs/1911.11875}
  {arXiv:1911.11875} \BibitemShut {NoStop}%
\bibitem [{\citenamefont {Furnstahl}\ \emph {et~al.}(2015)\citenamefont
  {Furnstahl}, \citenamefont {Klco}, \citenamefont {Phillips},\ and\
  \citenamefont {Wesolowski}}]{Furnstahl:2015rha}%
  \BibitemOpen
  \bibfield  {author} {\bibinfo {author} {\bibfnamefont {R.~J.}\ \bibnamefont
  {Furnstahl}}, \bibinfo {author} {\bibfnamefont {N.}~\bibnamefont {Klco}},
  \bibinfo {author} {\bibfnamefont {D.~R.}\ \bibnamefont {Phillips}},\ and\
  \bibinfo {author} {\bibfnamefont {S.}~\bibnamefont {Wesolowski}},\ }\bibfield
   {title} {\bibinfo {title} {{Quantifying truncation errors in effective field
  theory}},\ }\href {https://doi.org/10.1103/PhysRevC.92.024005} {\bibfield
  {journal} {\bibinfo  {journal} {Phys. Rev. C}\ }\textbf {\bibinfo {volume}
  {92}},\ \bibinfo {pages} {024005} (\bibinfo {year} {2015})},\ \Eprint
  {https://arxiv.org/abs/1506.01343} {arXiv:1506.01343} \BibitemShut {NoStop}%
%%CITATION = ARXIV:1506.01343;%%
\bibitem [{\citenamefont {Melendez}\ \emph {et~al.}(2017)\citenamefont
  {Melendez}, \citenamefont {Wesolowski},\ and\ \citenamefont
  {Furnstahl}}]{Melendez:2017phj}%
  \BibitemOpen
  \bibfield  {author} {\bibinfo {author} {\bibfnamefont {J.~A.}\ \bibnamefont
  {Melendez}}, \bibinfo {author} {\bibfnamefont {S.}~\bibnamefont
  {Wesolowski}},\ and\ \bibinfo {author} {\bibfnamefont {R.~J.}\ \bibnamefont
  {Furnstahl}},\ }\bibfield  {title} {\bibinfo {title} {{Bayesian truncation
  errors in chiral effective field theory: nucleon-nucleon observables}},\
  }\href {https://doi.org/10.1103/PhysRevC.96.024003} {\bibfield  {journal}
  {\bibinfo  {journal} {Phys. Rev. C}\ }\textbf {\bibinfo {volume} {96}},\
  \bibinfo {pages} {024003} (\bibinfo {year} {2017})},\ \Eprint
  {https://arxiv.org/abs/1704.03308} {arXiv:1704.03308} \BibitemShut {NoStop}%
%%CITATION = ARXIV:1704.03308;%%
\bibitem [{\citenamefont {Melendez}\ \emph {et~al.}(2019)\citenamefont
  {Melendez}, \citenamefont {Furnstahl}, \citenamefont {Phillips},
  \citenamefont {Pratola},\ and\ \citenamefont
  {Wesolowski}}]{Melendez:2019izc}%
  \BibitemOpen
  \bibfield  {author} {\bibinfo {author} {\bibfnamefont {J.~A.}\ \bibnamefont
  {Melendez}}, \bibinfo {author} {\bibfnamefont {R.~J.}\ \bibnamefont
  {Furnstahl}}, \bibinfo {author} {\bibfnamefont {D.~R.}\ \bibnamefont
  {Phillips}}, \bibinfo {author} {\bibfnamefont {M.~T.}\ \bibnamefont
  {Pratola}},\ and\ \bibinfo {author} {\bibfnamefont {S.}~\bibnamefont
  {Wesolowski}},\ }\bibfield  {title} {\bibinfo {title} {{Quantifying
  Correlated Truncation Errors in Effective Field Theory}},\ }\href
  {https://doi.org/10.1103/PhysRevC.100.044001} {\bibfield  {journal} {\bibinfo
   {journal} {Phys. Rev. C}\ }\textbf {\bibinfo {volume} {100}},\ \bibinfo
  {pages} {044001} (\bibinfo {year} {2019})},\ \Eprint
  {https://arxiv.org/abs/1904.10581} {arXiv:1904.10581} \BibitemShut {NoStop}%
%%CITATION = ARXIV:1904.10581;%%
\bibitem [{\citenamefont {Melendez}(2020)}]{Melendez:2020xcs}%
  \BibitemOpen
  \bibfield  {author} {\bibinfo {author} {\bibfnamefont {J.}~\bibnamefont
  {Melendez}},\ }\emph {\bibinfo {title} {{Effective Field Theory Truncation
  Errors and Why They Matter}}},\ \href
  {http://rave.ohiolink.edu/etdc/view?acc_num=osu1587114253866152} {Ph.D.
  thesis},\ \bibinfo  {school} {Ohio State U.} (\bibinfo {year}
  {2020})\BibitemShut {NoStop}%
\bibitem [{\citenamefont {Higdon}\ \emph {et~al.}(2015)\citenamefont {Higdon},
  \citenamefont {McDonnell}, \citenamefont {Schunck}, \citenamefont {Sarich},\
  and\ \citenamefont {Wild}}]{Higdon:2014tva}%
  \BibitemOpen
  \bibfield  {author} {\bibinfo {author} {\bibfnamefont {D.}~\bibnamefont
  {Higdon}}, \bibinfo {author} {\bibfnamefont {J.~D.}\ \bibnamefont
  {McDonnell}}, \bibinfo {author} {\bibfnamefont {N.}~\bibnamefont {Schunck}},
  \bibinfo {author} {\bibfnamefont {J.}~\bibnamefont {Sarich}},\ and\ \bibinfo
  {author} {\bibfnamefont {S.~M.}\ \bibnamefont {Wild}},\ }\bibfield  {title}
  {\bibinfo {title} {{A Bayesian Approach for Parameter Estimation and
  Prediction using a Computationally Intensive Model}},\ }\href
  {https://doi.org/10.1088/0954-3899/42/3/034009} {\bibfield  {journal}
  {\bibinfo  {journal} {J. Phys. G}\ }\textbf {\bibinfo {volume} {42}},\
  \bibinfo {pages} {034009} (\bibinfo {year} {2015})},\ \Eprint
  {https://arxiv.org/abs/1407.3017} {arXiv:1407.3017} \BibitemShut {NoStop}%
%%CITATION = ARXIV:1407.3017;%%
\bibitem [{\citenamefont {Wesolowski}\ \emph {et~al.}(2019)\citenamefont
  {Wesolowski}, \citenamefont {Furnstahl}, \citenamefont {Melendez},\ and\
  \citenamefont {Phillips}}]{Wesolowski:2018lzj}%
  \BibitemOpen
  \bibfield  {author} {\bibinfo {author} {\bibfnamefont {S.}~\bibnamefont
  {Wesolowski}}, \bibinfo {author} {\bibfnamefont {R.~J.}\ \bibnamefont
  {Furnstahl}}, \bibinfo {author} {\bibfnamefont {J.~A.}\ \bibnamefont
  {Melendez}},\ and\ \bibinfo {author} {\bibfnamefont {D.~R.}\ \bibnamefont
  {Phillips}},\ }\bibfield  {title} {\bibinfo {title} {{Exploring Bayesian
  parameter estimation for chiral effective field theory using
  nucleon–nucleon phase shifts}},\ }\href
  {https://doi.org/10.1088/1361-6471/aaf5fc} {\bibfield  {journal} {\bibinfo
  {journal} {J. Phys. G}\ }\textbf {\bibinfo {volume} {46}},\ \bibinfo {pages}
  {045102} (\bibinfo {year} {2019})},\ \Eprint
  {https://arxiv.org/abs/1808.08211} {arXiv:1808.08211} \BibitemShut {NoStop}%
%%CITATION = ARXIV:1808.08211;%%
\bibitem [{\citenamefont {Phillips}\ \emph {et~al.}(2021)\citenamefont
  {Phillips}, \citenamefont {Furnstahl}, \citenamefont {Heinz}, \citenamefont
  {Maiti}, \citenamefont {Nazarewicz}, \citenamefont {Nunes}, \citenamefont
  {Plumlee}, \citenamefont {Pratola}, \citenamefont {Pratt}, \citenamefont
  {Viens},\ and\ \citenamefont {Wild}}]{Phillips:2020dmw}%
  \BibitemOpen
  \bibfield  {author} {\bibinfo {author} {\bibfnamefont {D.~R.}\ \bibnamefont
  {Phillips}}, \bibinfo {author} {\bibfnamefont {R.~J.}\ \bibnamefont
  {Furnstahl}}, \bibinfo {author} {\bibfnamefont {U.}~\bibnamefont {Heinz}},
  \bibinfo {author} {\bibfnamefont {T.}~\bibnamefont {Maiti}}, \bibinfo
  {author} {\bibfnamefont {W.}~\bibnamefont {Nazarewicz}}, \bibinfo {author}
  {\bibfnamefont {F.~M.}\ \bibnamefont {Nunes}}, \bibinfo {author}
  {\bibfnamefont {M.}~\bibnamefont {Plumlee}}, \bibinfo {author} {\bibfnamefont
  {M.~T.}\ \bibnamefont {Pratola}}, \bibinfo {author} {\bibfnamefont
  {S.}~\bibnamefont {Pratt}}, \bibinfo {author} {\bibfnamefont {F.~G.}\
  \bibnamefont {Viens}},\ and\ \bibinfo {author} {\bibfnamefont {S.~M.}\
  \bibnamefont {Wild}},\ }\bibfield  {title} {\bibinfo {title} {{Get on the
  BAND Wagon: A Bayesian Framework for Quantifying Model Uncertainties in
  Nuclear Dynamics}},\ }\href {https://doi.org/10.1088/1361-6471/abf1df}
  {\bibfield  {journal} {\bibinfo  {journal} {J. Phys. G}\ }\textbf {\bibinfo
  {volume} {48}},\ \bibinfo {pages} {072001} (\bibinfo {year} {2021})},\
  \Eprint {https://arxiv.org/abs/2012.07704} {arXiv:2012.07704 [nucl-th]}
  \BibitemShut {NoStop}%
\bibitem [{\citenamefont {Melendez}\ \emph
  {et~al.}(2021{\natexlab{a}})\citenamefont {Melendez}, \citenamefont
  {Furnstahl}, \citenamefont {Grie{\ss}hammer}, \citenamefont {McGovern},
  \citenamefont {Phillips},\ and\ \citenamefont {Pratola}}]{Melendez:2020ikd}%
  \BibitemOpen
  \bibfield  {author} {\bibinfo {author} {\bibfnamefont {J.~A.}\ \bibnamefont
  {Melendez}}, \bibinfo {author} {\bibfnamefont {R.~J.}\ \bibnamefont
  {Furnstahl}}, \bibinfo {author} {\bibfnamefont {H.~W.}\ \bibnamefont
  {Grie{\ss}hammer}}, \bibinfo {author} {\bibfnamefont {J.~A.}\ \bibnamefont
  {McGovern}}, \bibinfo {author} {\bibfnamefont {D.~R.}\ \bibnamefont
  {Phillips}},\ and\ \bibinfo {author} {\bibfnamefont {M.~T.}\ \bibnamefont
  {Pratola}},\ }\bibfield  {title} {\bibinfo {title} {{Designing Optimal
  Experiments: An Application to Proton Compton Scattering}},\ }\href
  {https://doi.org/10.1140/epja/s10050-021-00382-2} {\bibfield  {journal}
  {\bibinfo  {journal} {Eur. Phys. J. A}\ }\textbf {\bibinfo {volume} {57}},\
  \bibinfo {pages} {81} (\bibinfo {year} {2021}{\natexlab{a}})},\ \Eprint
  {https://arxiv.org/abs/2004.11307} {arXiv:2004.11307 [nucl-th]} \BibitemShut
  {NoStop}%
%%CITATION = ARXIV:2004.11307;%%
\bibitem [{\citenamefont {Svensson}\ \emph {et~al.}(2022)\citenamefont
  {Svensson}, \citenamefont {Ekstr\"om},\ and\ \citenamefont
  {Forss\'en}}]{Svensson:2022kkj}%
  \BibitemOpen
  \bibfield  {author} {\bibinfo {author} {\bibfnamefont {I.}~\bibnamefont
  {Svensson}}, \bibinfo {author} {\bibfnamefont {A.}~\bibnamefont
  {Ekstr\"om}},\ and\ \bibinfo {author} {\bibfnamefont {C.}~\bibnamefont
  {Forss\'en}},\ }\bibfield  {title} {\bibinfo {title} {{Bayesian estimation of
  the low-energy constants up to fourth order in the nucleon-nucleon sector of
  chiral effective field theory}},\ }\href@noop {} {\  (\bibinfo {year}
  {2022})},\ \Eprint {https://arxiv.org/abs/2206.08250} {arXiv:2206.08250}
  \BibitemShut {NoStop}%
\bibitem [{\citenamefont {Drischler}\ and\ \citenamefont
  {Zhang}(2022)}]{Drischler:2022yfb}%
  \BibitemOpen
  \bibfield  {author} {\bibinfo {author} {\bibfnamefont {C.}~\bibnamefont
  {Drischler}}\ and\ \bibinfo {author} {\bibfnamefont {X.}~\bibnamefont
  {Zhang}},\ }\bibfield  {title} {\bibinfo {title} {Few-body emulators based on
  eigenvector continuation},\ }in\ \href
  {https://doi.org/10.1007/s00601-022-01749-x} {\emph {\bibinfo {booktitle}
  {{Nuclear Forces for Precision Nuclear Physics: A Collection of
  Perspectives}}}},\ Vol.~\bibinfo {volume} {63},\ \bibinfo {editor} {edited
  by\ \bibinfo {editor} {\bibfnamefont {I.}~\bibnamefont {Tews}}, \bibinfo
  {editor} {\bibfnamefont {Z.}~\bibnamefont {Davoudi}}, \bibinfo {editor}
  {\bibfnamefont {A.}~\bibnamefont {Ekstr{\"o}m}},\ and\ \bibinfo {editor}
  {\bibfnamefont {J.~D.}\ \bibnamefont {Holt}}}\ (\bibinfo {year} {2022})\
  Chap.~\bibinfo {chapter} {8}, p.~\bibinfo {pages} {67},\ \Eprint
  {https://arxiv.org/abs/2202.01105} {arXiv:2202.01105} \BibitemShut {NoStop}%
\bibitem [{\citenamefont {Furnstahl}\ \emph {et~al.}(2020)\citenamefont
  {Furnstahl}, \citenamefont {Garcia}, \citenamefont {Millican},\ and\
  \citenamefont {Zhang}}]{Furnstahl:2020abp}%
  \BibitemOpen
  \bibfield  {author} {\bibinfo {author} {\bibfnamefont {R.~J.}\ \bibnamefont
  {Furnstahl}}, \bibinfo {author} {\bibfnamefont {A.~J.}\ \bibnamefont
  {Garcia}}, \bibinfo {author} {\bibfnamefont {P.~J.}\ \bibnamefont
  {Millican}},\ and\ \bibinfo {author} {\bibfnamefont {X.}~\bibnamefont
  {Zhang}},\ }\bibfield  {title} {\bibinfo {title} {{Efficient emulators for
  scattering using eigenvector continuation}},\ }\href
  {https://doi.org/10.1016/j.physletb.2020.135719} {\bibfield  {journal}
  {\bibinfo  {journal} {Phys. Lett. B}\ }\textbf {\bibinfo {volume} {809}},\
  \bibinfo {pages} {135719} (\bibinfo {year} {2020})},\ \Eprint
  {https://arxiv.org/abs/2007.03635} {arXiv:2007.03635 [nucl-th]} \BibitemShut
  {NoStop}%
\bibitem [{\citenamefont {Drischler}\ \emph {et~al.}(2021)\citenamefont
  {Drischler}, \citenamefont {Quinonez}, \citenamefont {Giuliani},
  \citenamefont {Lovell},\ and\ \citenamefont {Nunes}}]{Drischler:2021qoy}%
  \BibitemOpen
  \bibfield  {author} {\bibinfo {author} {\bibfnamefont {C.}~\bibnamefont
  {Drischler}}, \bibinfo {author} {\bibfnamefont {M.}~\bibnamefont {Quinonez}},
  \bibinfo {author} {\bibfnamefont {P.~G.}\ \bibnamefont {Giuliani}}, \bibinfo
  {author} {\bibfnamefont {A.~E.}\ \bibnamefont {Lovell}},\ and\ \bibinfo
  {author} {\bibfnamefont {F.~M.}\ \bibnamefont {Nunes}},\ }\bibfield  {title}
  {\bibinfo {title} {{Toward emulating nuclear reactions using eigenvector
  continuation}},\ }\href {https://doi.org/10.1016/j.physletb.2021.136777}
  {\bibfield  {journal} {\bibinfo  {journal} {Phys. Lett. B}\ }\textbf
  {\bibinfo {volume} {823}},\ \bibinfo {pages} {136777} (\bibinfo {year}
  {2021})},\ \Eprint {https://arxiv.org/abs/2108.08269} {arXiv:2108.08269
  [nucl-th]} \BibitemShut {NoStop}%
\bibitem [{\citenamefont {Melendez}\ \emph
  {et~al.}(2021{\natexlab{b}})\citenamefont {Melendez}, \citenamefont
  {Drischler}, \citenamefont {Garcia}, \citenamefont {Furnstahl},\ and\
  \citenamefont {Zhang}}]{Melendez:2021lyq}%
  \BibitemOpen
  \bibfield  {author} {\bibinfo {author} {\bibfnamefont {J.}~\bibnamefont
  {Melendez}}, \bibinfo {author} {\bibfnamefont {C.}~\bibnamefont {Drischler}},
  \bibinfo {author} {\bibfnamefont {A.}~\bibnamefont {Garcia}}, \bibinfo
  {author} {\bibfnamefont {R.}~\bibnamefont {Furnstahl}},\ and\ \bibinfo
  {author} {\bibfnamefont {X.}~\bibnamefont {Zhang}},\ }\bibfield  {title}
  {\bibinfo {title} {Fast \& accurate emulation of two-body scattering
  observables without wave functions},\ }\href
  {https://doi.org/https://doi.org/10.1016/j.physletb.2021.136608} {\bibfield
  {journal} {\bibinfo  {journal} {Phys. Lett. B}\ }\textbf {\bibinfo {volume}
  {821}},\ \bibinfo {pages} {136608} (\bibinfo {year}
  {2021}{\natexlab{b}})}\BibitemShut {NoStop}%
\bibitem [{\citenamefont {Bastos}\ and\ \citenamefont
  {O'Hagan}(2009)}]{BastosDiagnosticsGaussianProcess2009}%
  \BibitemOpen
  \bibfield  {author} {\bibinfo {author} {\bibfnamefont {L.~S.}\ \bibnamefont
  {Bastos}}\ and\ \bibinfo {author} {\bibfnamefont {A.}~\bibnamefont
  {O'Hagan}},\ }\bibfield  {title} {\bibinfo {title} {Diagnostics for
  {{Gaussian Process Emulators}}},\ }\href
  {https://doi.org/10.1198/TECH.2009.08019} {\bibfield  {journal} {\bibinfo
  {journal} {Technometrics}\ }\textbf {\bibinfo {volume} {51}},\ \bibinfo
  {pages} {425} (\bibinfo {year} {2009})}\BibitemShut {NoStop}%
\bibitem [{\citenamefont {Melendez}\ \emph {et~al.}(2022)\citenamefont
  {Melendez}, \citenamefont {Drischler}, \citenamefont {Furnstahl},
  \citenamefont {Garcia},\ and\ \citenamefont {Zhang}}]{Melendez:2022kid}%
  \BibitemOpen
  \bibfield  {author} {\bibinfo {author} {\bibfnamefont {J.~A.}\ \bibnamefont
  {Melendez}}, \bibinfo {author} {\bibfnamefont {C.}~\bibnamefont {Drischler}},
  \bibinfo {author} {\bibfnamefont {R.~J.}\ \bibnamefont {Furnstahl}}, \bibinfo
  {author} {\bibfnamefont {A.~J.}\ \bibnamefont {Garcia}},\ and\ \bibinfo
  {author} {\bibfnamefont {X.}~\bibnamefont {Zhang}},\ }\bibfield  {title}
  {\bibinfo {title} {{Model reduction methods for nuclear emulators}},\ }\href
  {https://doi.org/10.1088/1361-6471/ac83dd} {\bibfield  {journal} {\bibinfo
  {journal} {J. Phys. G}\ }\textbf {\bibinfo {volume} {49}},\ \bibinfo {pages}
  {102001} (\bibinfo {year} {2022})},\ \Eprint
  {https://arxiv.org/abs/2203.05528} {arXiv:2203.05528 [nucl-th]} \BibitemShut
  {NoStop}%
\bibitem [{\citenamefont {Bonilla}\ \emph {et~al.}(2022)\citenamefont
  {Bonilla}, \citenamefont {Giuliani}, \citenamefont {Godbey},\ and\
  \citenamefont {Lee}}]{Bonilla:2022rph}%
  \BibitemOpen
  \bibfield  {author} {\bibinfo {author} {\bibfnamefont {E.}~\bibnamefont
  {Bonilla}}, \bibinfo {author} {\bibfnamefont {P.}~\bibnamefont {Giuliani}},
  \bibinfo {author} {\bibfnamefont {K.}~\bibnamefont {Godbey}},\ and\ \bibinfo
  {author} {\bibfnamefont {D.}~\bibnamefont {Lee}},\ }\bibfield  {title}
  {\bibinfo {title} {{Training and projecting: A reduced basis method emulator
  for many-body physics}},\ }\href
  {https://doi.org/10.1103/PhysRevC.106.054322} {\bibfield  {journal} {\bibinfo
   {journal} {Phys. Rev. C}\ }\textbf {\bibinfo {volume} {106}},\ \bibinfo
  {pages} {054322} (\bibinfo {year} {2022})},\ \Eprint
  {https://arxiv.org/abs/2203.05284} {arXiv:2203.05284 [nucl-th]} \BibitemShut
  {NoStop}%
\bibitem [{\citenamefont {Drischler}\ \emph {et~al.}(2023)\citenamefont
  {Drischler}, \citenamefont {Melendez}, \citenamefont {Furnstahl},
  \citenamefont {Garcia},\ and\ \citenamefont {Zhang}}]{Drischler:2022ipa}%
  \BibitemOpen
  \bibfield  {author} {\bibinfo {author} {\bibfnamefont {C.}~\bibnamefont
  {Drischler}}, \bibinfo {author} {\bibfnamefont {J.~A.}\ \bibnamefont
  {Melendez}}, \bibinfo {author} {\bibfnamefont {R.~J.}\ \bibnamefont
  {Furnstahl}}, \bibinfo {author} {\bibfnamefont {A.~J.}\ \bibnamefont
  {Garcia}},\ and\ \bibinfo {author} {\bibfnamefont {X.}~\bibnamefont
  {Zhang}},\ }\bibfield  {title} {\bibinfo {title} {{BUQEYE Guide to
  Projection-Based Emulators in Nuclear Physics}},\ }\href
  {https://doi.org/10.3389/fphy.2022.1092931} {\bibfield  {journal} {\bibinfo
  {journal} {Front. Phys.}\ }\textbf {\bibinfo {volume} {10}},\ \bibinfo
  {pages} {92931} (\bibinfo {year} {2023})},\ \bibinfo {note} {supplemental,
  interactive Python code can be found on the companion
  website~\url{https://github.com/buqeye/frontiers-emulator-review}},\ \Eprint
  {https://arxiv.org/abs/2212.04912} {arXiv:2212.04912} \BibitemShut {NoStop}%
\bibitem [{\citenamefont {Frame}\ \emph {et~al.}(2018)\citenamefont {Frame},
  \citenamefont {He}, \citenamefont {Ipsen}, \citenamefont {Lee}, \citenamefont
  {Lee},\ and\ \citenamefont {Rrapaj}}]{Frame:2017fah}%
  \BibitemOpen
  \bibfield  {author} {\bibinfo {author} {\bibfnamefont {D.}~\bibnamefont
  {Frame}}, \bibinfo {author} {\bibfnamefont {R.}~\bibnamefont {He}}, \bibinfo
  {author} {\bibfnamefont {I.}~\bibnamefont {Ipsen}}, \bibinfo {author}
  {\bibfnamefont {D.}~\bibnamefont {Lee}}, \bibinfo {author} {\bibfnamefont
  {D.}~\bibnamefont {Lee}},\ and\ \bibinfo {author} {\bibfnamefont
  {E.}~\bibnamefont {Rrapaj}},\ }\bibfield  {title} {\bibinfo {title}
  {{Eigenvector continuation with subspace learning}},\ }\href
  {https://doi.org/10.1103/PhysRevLett.121.032501} {\bibfield  {journal}
  {\bibinfo  {journal} {Phys. Rev. Lett.}\ }\textbf {\bibinfo {volume} {121}},\
  \bibinfo {pages} {032501} (\bibinfo {year} {2018})},\ \Eprint
  {https://arxiv.org/abs/1711.07090} {arXiv:1711.07090} \BibitemShut {NoStop}%
%%CITATION = ARXIV:1711.07090;%%
\bibitem [{\citenamefont {K\"onig}\ \emph {et~al.}(2020)\citenamefont
  {K\"onig}, \citenamefont {Ekstr\"om}, \citenamefont {Hebeler}, \citenamefont
  {Lee},\ and\ \citenamefont {Schwenk}}]{Konig:2019adq}%
  \BibitemOpen
  \bibfield  {author} {\bibinfo {author} {\bibfnamefont {S.}~\bibnamefont
  {K\"onig}}, \bibinfo {author} {\bibfnamefont {A.}~\bibnamefont {Ekstr\"om}},
  \bibinfo {author} {\bibfnamefont {K.}~\bibnamefont {Hebeler}}, \bibinfo
  {author} {\bibfnamefont {D.}~\bibnamefont {Lee}},\ and\ \bibinfo {author}
  {\bibfnamefont {A.}~\bibnamefont {Schwenk}},\ }\bibfield  {title} {\bibinfo
  {title} {{Eigenvector Continuation as an Efficient and Accurate Emulator for
  Uncertainty Quantification}},\ }\href
  {https://doi.org/10.1016/j.physletb.2020.135814} {\bibfield  {journal}
  {\bibinfo  {journal} {Phys. Lett. B}\ }\textbf {\bibinfo {volume} {810}},\
  \bibinfo {pages} {135814} (\bibinfo {year} {2020})},\ \Eprint
  {https://arxiv.org/abs/1909.08446} {arXiv:1909.08446 [nucl-th]} \BibitemShut
  {NoStop}%
\bibitem [{\citenamefont {Kohn}(1948)}]{Kohn:1948col}%
  \BibitemOpen
  \bibfield  {author} {\bibinfo {author} {\bibfnamefont {W.}~\bibnamefont
  {Kohn}},\ }\bibfield  {title} {\bibinfo {title} {{Variational Methods in
  Nuclear Collision Problems}},\ }\href
  {https://doi.org/10.1103/PhysRev.74.1763} {\bibfield  {journal} {\bibinfo
  {journal} {Phys. Rev.}\ }\textbf {\bibinfo {volume} {74}},\ \bibinfo {pages}
  {1763} (\bibinfo {year} {1948})}\BibitemShut {NoStop}%
\bibitem [{\citenamefont {Taylor}(2006)}]{taylor2006scattering}%
  \BibitemOpen
  \bibfield  {author} {\bibinfo {author} {\bibfnamefont {J.~R.}\ \bibnamefont
  {Taylor}},\ }\href@noop {} {\emph {\bibinfo {title} {Scattering Theory: The
  Quantum Theory of Nonrelativistic Collisions}}}\ (\bibinfo  {publisher}
  {Dover},\ \bibinfo {address} {Mineola, New York},\ \bibinfo {year}
  {2006})\BibitemShut {NoStop}%
\bibitem [{\citenamefont {Zhang}\ and\ \citenamefont
  {Furnstahl}(2022)}]{Zhang:2021jmi}%
  \BibitemOpen
  \bibfield  {author} {\bibinfo {author} {\bibfnamefont {X.}~\bibnamefont
  {Zhang}}\ and\ \bibinfo {author} {\bibfnamefont {R.~J.}\ \bibnamefont
  {Furnstahl}},\ }\bibfield  {title} {\bibinfo {title} {Fast emulation of
  quantum three-body scattering},\ }\href
  {https://doi.org/10.1103/PhysRevC.105.064004} {\bibfield  {journal} {\bibinfo
   {journal} {Phys. Rev. C}\ }\textbf {\bibinfo {volume} {105}},\ \bibinfo
  {pages} {064004} (\bibinfo {year} {2022})},\ \Eprint
  {https://arxiv.org/abs/2110.04269} {arXiv:2110.04269 [nucl-th]} \BibitemShut
  {NoStop}%
\bibitem [{\citenamefont {Newton}(2002)}]{newton2002scattering}%
  \BibitemOpen
  \bibfield  {author} {\bibinfo {author} {\bibfnamefont {R.~G.}\ \bibnamefont
  {Newton}},\ }\href@noop {} {\emph {\bibinfo {title} {Scattering theory of
  waves and particles}}}\ (\bibinfo  {publisher} {Dover},\ \bibinfo {address}
  {Mineola, New York},\ \bibinfo {year} {2002})\BibitemShut {NoStop}%
\bibitem [{\citenamefont {Schwartz}(1961)}]{PhysRev.124.1468}%
  \BibitemOpen
  \bibfield  {author} {\bibinfo {author} {\bibfnamefont {C.}~\bibnamefont
  {Schwartz}},\ }\bibfield  {title} {\bibinfo {title} {Electron scattering from
  hydrogen},\ }\href {https://doi.org/10.1103/PhysRev.124.1468} {\bibfield
  {journal} {\bibinfo  {journal} {Phys. Rev.}\ }\textbf {\bibinfo {volume}
  {124}},\ \bibinfo {pages} {1468} (\bibinfo {year} {1961})}\BibitemShut
  {NoStop}%
\bibitem [{\citenamefont {Nesbet}(1980)}]{nesbet1980variational}%
  \BibitemOpen
  \bibfield  {author} {\bibinfo {author} {\bibfnamefont {R.}~\bibnamefont
  {Nesbet}},\ }\href {https://books.google.com/books?id=0mF5AAAAIAAJ} {\emph
  {\bibinfo {title} {Variational methods in electron-atom scattering
  theory}}},\ Physics of atoms and molecules\ (\bibinfo  {publisher} {Plenum
  Press},\ \bibinfo {year} {1980})\BibitemShut {NoStop}%
\bibitem [{\citenamefont {{{BUQEYE collaboration}}}(2022)}]{BUQEYEsoftware}%
  \BibitemOpen
  \bibfield  {author} {\bibinfo {author} {\bibnamefont {{{BUQEYE
  collaboration}}}},\ }\href {\mbox{https://buqeye.github.io/software/}} {}
  (\bibinfo {year} {2022}),\ \bibinfo {note}
  {\url{https://buqeye.github.io/software/}}\BibitemShut {NoStop}%
\bibitem [{\citenamefont {Lucchese}(1989)}]{Lucchese:1989zz}%
  \BibitemOpen
  \bibfield  {author} {\bibinfo {author} {\bibfnamefont {R.~R.}\ \bibnamefont
  {Lucchese}},\ }\bibfield  {title} {\bibinfo {title} {Anomalous singularities
  in the complex kohn variational principle of quantum scattering theory},\
  }\href {https://doi.org/10.1103/PhysRevA.40.6879} {\bibfield  {journal}
  {\bibinfo  {journal} {Phys. Rev. A}\ }\textbf {\bibinfo {volume} {40}},\
  \bibinfo {pages} {6879} (\bibinfo {year} {1989})}\BibitemShut {NoStop}%
\bibitem [{\citenamefont {Morrison}\ and\ \citenamefont
  {Feldt}(2007)}]{Morrison:2007}%
  \BibitemOpen
  \bibfield  {author} {\bibinfo {author} {\bibfnamefont {M.~A.}\ \bibnamefont
  {Morrison}}\ and\ \bibinfo {author} {\bibfnamefont {A.~N.}\ \bibnamefont
  {Feldt}},\ }\bibfield  {title} {\bibinfo {title} {{Through scattering theory
  with gun and camera: Coping with conventions in collision theory}},\
  }\bibfield  {journal} {\bibinfo  {journal} {Am. J. Phys.}\ }\textbf {\bibinfo
  {volume} {75}},\ \href {https://doi.org/https://doi.org/10.1119/1.2358156}
  {https://doi.org/10.1119/1.2358156} (\bibinfo {year} {2007})\BibitemShut
  {NoStop}%
\bibitem [{\citenamefont {Kamimura}(1977)}]{10.1143/PTPS.62.236}%
  \BibitemOpen
  \bibfield  {author} {\bibinfo {author} {\bibfnamefont {M.}~\bibnamefont
  {Kamimura}},\ }\bibfield  {title} {\bibinfo {title} {{Chapter V. A Coupled
  Channel Variational Method for Microscopic Study of Reactions between Complex
  Nuclei}},\ }\href {https://doi.org/10.1143/PTPS.62.236} {\bibfield  {journal}
  {\bibinfo  {journal} {Progress of Theoretical Physics Supplement}\ }\textbf
  {\bibinfo {volume} {62}},\ \bibinfo {pages} {236} (\bibinfo {year}
  {1977})}\BibitemShut {NoStop}%
\bibitem [{\citenamefont {Stoks}\ and\ \citenamefont
  {de~Swart}(1990)}]{Stoks:1990us}%
  \BibitemOpen
  \bibfield  {author} {\bibinfo {author} {\bibfnamefont {V.~G.~J.}\
  \bibnamefont {Stoks}}\ and\ \bibinfo {author} {\bibfnamefont {J.~J.}\
  \bibnamefont {de~Swart}},\ }\bibfield  {title} {\bibinfo {title} {{The
  Magnetic moment interaction in nucleon-nucleon phase shift analyses}},\
  }\href {https://doi.org/10.1103/PhysRevC.42.1235} {\bibfield  {journal}
  {\bibinfo  {journal} {Phys. Rev.}\ }\textbf {\bibinfo {volume} {C42}},\
  \bibinfo {pages} {1235} (\bibinfo {year} {1990})}\BibitemShut {NoStop}%
%%CITATION = PHRVA,C42,1235;%%
\bibitem [{\citenamefont {Haftel}\ and\ \citenamefont
  {Tabakin}(1970)}]{Haftel:1970zz}%
  \BibitemOpen
  \bibfield  {author} {\bibinfo {author} {\bibfnamefont {M.~I.}\ \bibnamefont
  {Haftel}}\ and\ \bibinfo {author} {\bibfnamefont {F.}~\bibnamefont
  {Tabakin}},\ }\bibfield  {title} {\bibinfo {title} {{Nuclear saturation and
  the smoothness of nucleon-nucleon potentials}},\ }\href
  {https://doi.org/10.1016/0375-9474(70)90047-3} {\bibfield  {journal}
  {\bibinfo  {journal} {Nucl. Phys. A}\ }\textbf {\bibinfo {volume} {158}},\
  \bibinfo {pages} {1} (\bibinfo {year} {1970})}\BibitemShut {NoStop}%
\bibitem [{\citenamefont {Reinert}\ \emph {et~al.}(2018)\citenamefont
  {Reinert}, \citenamefont {Krebs},\ and\ \citenamefont
  {Epelbaum}}]{Reinert:2017usi}%
  \BibitemOpen
  \bibfield  {author} {\bibinfo {author} {\bibfnamefont {P.}~\bibnamefont
  {Reinert}}, \bibinfo {author} {\bibfnamefont {H.}~\bibnamefont {Krebs}},\
  and\ \bibinfo {author} {\bibfnamefont {E.}~\bibnamefont {Epelbaum}},\
  }\bibfield  {title} {\bibinfo {title} {{Semilocal momentum-space regularized
  chiral two-nucleon potentials up to fifth order}},\ }\href
  {https://doi.org/10.1140/epja/i2018-12516-4} {\bibfield  {journal} {\bibinfo
  {journal} {Eur. Phys. J. A}\ }\textbf {\bibinfo {volume} {54}},\ \bibinfo
  {pages} {86} (\bibinfo {year} {2018})},\ \Eprint
  {https://arxiv.org/abs/1711.08821} {arXiv:1711.08821} \BibitemShut {NoStop}%
%%CITATION = ARXIV:1711.08821;%%
\bibitem [{\citenamefont {Landau}(1996)}]{Landau:1996}%
  \BibitemOpen
  \bibfield  {author} {\bibinfo {author} {\bibfnamefont {R.~H.}\ \bibnamefont
  {Landau}},\ }\href@noop {} {\emph {\bibinfo {title} {Quantum Mechanics
  II}}},\ \bibinfo {edition} {2nd}\ ed.\ (\bibinfo  {publisher} {John Wiley \&
  Sons, Inc.},\ \bibinfo {address} {New York},\ \bibinfo {year}
  {1996})\BibitemShut {NoStop}%
\bibitem [{\citenamefont {Gl{\"o}ckle}\ \emph {et~al.}(1982)\citenamefont
  {Gl{\"o}ckle}, \citenamefont {Hasberg},\ and\ \citenamefont
  {Neghabian}}]{GlockleInterpolation1982}%
  \BibitemOpen
  \bibfield  {author} {\bibinfo {author} {\bibfnamefont {W.}~\bibnamefont
  {Gl{\"o}ckle}}, \bibinfo {author} {\bibfnamefont {G.}~\bibnamefont
  {Hasberg}},\ and\ \bibinfo {author} {\bibfnamefont {A.~R.}\ \bibnamefont
  {Neghabian}},\ }\bibfield  {title} {\bibinfo {title} {Numerical treatment of
  few body equations in momentum space by the spline method},\ }\href
  {https://doi.org/10.1007/BF01417437} {\bibfield  {journal} {\bibinfo
  {journal} {Z. Phys. A-Hadron Nucl.}\ }\textbf {\bibinfo {volume} {305}},\
  \bibinfo {pages} {217} (\bibinfo {year} {1982})}\BibitemShut {NoStop}%
\bibitem [{\citenamefont {Harris}\ \emph {et~al.}(2020)\citenamefont {Harris},
  \citenamefont {Millman}, \citenamefont {van~der Walt}, \citenamefont
  {Gommers}, \citenamefont {Virtanen}, \citenamefont {Cournapeau},
  \citenamefont {Wieser}, \citenamefont {Taylor}, \citenamefont {Berg},
  \citenamefont {Smith}, \citenamefont {Kern}, \citenamefont {Picus},
  \citenamefont {Hoyer}, \citenamefont {van Kerkwijk}, \citenamefont {Brett},
  \citenamefont {Haldane}, \citenamefont {del R{\'{i}}o}, \citenamefont
  {Wiebe}, \citenamefont {Peterson}, \citenamefont {G{\'{e}}rard-Marchant},
  \citenamefont {Sheppard}, \citenamefont {Reddy}, \citenamefont {Weckesser},
  \citenamefont {Abbasi}, \citenamefont {Gohlke},\ and\ \citenamefont
  {Oliphant}}]{harris2020array}%
  \BibitemOpen
  \bibfield  {author} {\bibinfo {author} {\bibfnamefont {C.~R.}\ \bibnamefont
  {Harris}}, \bibinfo {author} {\bibfnamefont {K.~J.}\ \bibnamefont {Millman}},
  \bibinfo {author} {\bibfnamefont {S.~J.}\ \bibnamefont {van~der Walt}},
  \bibinfo {author} {\bibfnamefont {R.}~\bibnamefont {Gommers}}, \bibinfo
  {author} {\bibfnamefont {P.}~\bibnamefont {Virtanen}}, \bibinfo {author}
  {\bibfnamefont {D.}~\bibnamefont {Cournapeau}}, \bibinfo {author}
  {\bibfnamefont {E.}~\bibnamefont {Wieser}}, \bibinfo {author} {\bibfnamefont
  {J.}~\bibnamefont {Taylor}}, \bibinfo {author} {\bibfnamefont
  {S.}~\bibnamefont {Berg}}, \bibinfo {author} {\bibfnamefont {N.~J.}\
  \bibnamefont {Smith}}, \bibinfo {author} {\bibfnamefont {R.}~\bibnamefont
  {Kern}}, \bibinfo {author} {\bibfnamefont {M.}~\bibnamefont {Picus}},
  \bibinfo {author} {\bibfnamefont {S.}~\bibnamefont {Hoyer}}, \bibinfo
  {author} {\bibfnamefont {M.~H.}\ \bibnamefont {van Kerkwijk}}, \bibinfo
  {author} {\bibfnamefont {M.}~\bibnamefont {Brett}}, \bibinfo {author}
  {\bibfnamefont {A.}~\bibnamefont {Haldane}}, \bibinfo {author} {\bibfnamefont
  {J.~F.}\ \bibnamefont {del R{\'{i}}o}}, \bibinfo {author} {\bibfnamefont
  {M.}~\bibnamefont {Wiebe}}, \bibinfo {author} {\bibfnamefont
  {P.}~\bibnamefont {Peterson}}, \bibinfo {author} {\bibfnamefont
  {P.}~\bibnamefont {G{\'{e}}rard-Marchant}}, \bibinfo {author} {\bibfnamefont
  {K.}~\bibnamefont {Sheppard}}, \bibinfo {author} {\bibfnamefont
  {T.}~\bibnamefont {Reddy}}, \bibinfo {author} {\bibfnamefont
  {W.}~\bibnamefont {Weckesser}}, \bibinfo {author} {\bibfnamefont
  {H.}~\bibnamefont {Abbasi}}, \bibinfo {author} {\bibfnamefont
  {C.}~\bibnamefont {Gohlke}},\ and\ \bibinfo {author} {\bibfnamefont {T.~E.}\
  \bibnamefont {Oliphant}},\ }\bibfield  {title} {\bibinfo {title} {Array
  programming with {NumPy}},\ }\href
  {https://doi.org/10.1038/s41586-020-2649-2} {\bibfield  {journal} {\bibinfo
  {journal} {Nature}\ }\textbf {\bibinfo {volume} {585}},\ \bibinfo {pages}
  {357} (\bibinfo {year} {2020})}\BibitemShut {NoStop}%
\bibitem [{\citenamefont {Viviani}\ \emph {et~al.}(2001)\citenamefont
  {Viviani}, \citenamefont {Kievsky},\ and\ \citenamefont
  {Rosati}}]{Viviani:2001sy}%
  \BibitemOpen
  \bibfield  {author} {\bibinfo {author} {\bibfnamefont {M.}~\bibnamefont
  {Viviani}}, \bibinfo {author} {\bibfnamefont {A.}~\bibnamefont {Kievsky}},\
  and\ \bibinfo {author} {\bibfnamefont {S.}~\bibnamefont {Rosati}},\
  }\bibfield  {title} {\bibinfo {title} {{The Kohn variational principle for
  elastic proton deuteron scattering above deuteron breakup threshold}},\
  }\href {https://doi.org/10.1007/s006010170017} {\bibfield  {journal}
  {\bibinfo  {journal} {Few Body Syst.}\ }\textbf {\bibinfo {volume} {30}},\
  \bibinfo {pages} {39} (\bibinfo {year} {2001})},\ \Eprint
  {https://arxiv.org/abs/nucl-th/0102048} {arXiv:nucl-th/0102048} \BibitemShut
  {NoStop}%
\bibitem [{\citenamefont {Carlsson}\ \emph {et~al.}(2016)\citenamefont
  {Carlsson}, \citenamefont {Ekström}, \citenamefont {Forssén}, \citenamefont
  {Strömberg}, \citenamefont {Jansen}, \citenamefont {Lilja}, \citenamefont
  {Lindby}, \citenamefont {Mattsson},\ and\ \citenamefont
  {Wendt}}]{Carlsson:2015vda}%
  \BibitemOpen
  \bibfield  {author} {\bibinfo {author} {\bibfnamefont {B.~D.}\ \bibnamefont
  {Carlsson}}, \bibinfo {author} {\bibfnamefont {A.}~\bibnamefont {Ekström}},
  \bibinfo {author} {\bibfnamefont {C.}~\bibnamefont {Forssén}}, \bibinfo
  {author} {\bibfnamefont {D.~F.}\ \bibnamefont {Strömberg}}, \bibinfo
  {author} {\bibfnamefont {G.~R.}\ \bibnamefont {Jansen}}, \bibinfo {author}
  {\bibfnamefont {O.}~\bibnamefont {Lilja}}, \bibinfo {author} {\bibfnamefont
  {M.}~\bibnamefont {Lindby}}, \bibinfo {author} {\bibfnamefont {B.~A.}\
  \bibnamefont {Mattsson}},\ and\ \bibinfo {author} {\bibfnamefont {K.~A.}\
  \bibnamefont {Wendt}},\ }\bibfield  {title} {\bibinfo {title} {{Uncertainty
  analysis and order-by-order optimization of chiral nuclear interactions}},\
  }\href {https://doi.org/10.1103/PhysRevX.6.011019} {\bibfield  {journal}
  {\bibinfo  {journal} {Phys. Rev. X}\ }\textbf {\bibinfo {volume} {6}},\
  \bibinfo {pages} {011019} (\bibinfo {year} {2016})},\ \Eprint
  {https://arxiv.org/abs/1506.02466} {arXiv:1506.02466} \BibitemShut {NoStop}%
%%CITATION = ARXIV:1506.02466;%%
\bibitem [{\citenamefont {Stoks}\ \emph {et~al.}(1993)\citenamefont {Stoks},
  \citenamefont {Klomp}, \citenamefont {Rentmeester},\ and\ \citenamefont
  {de~Swart}}]{Stoks:1993tb}%
  \BibitemOpen
  \bibfield  {author} {\bibinfo {author} {\bibfnamefont {V.~G.~J.}\
  \bibnamefont {Stoks}}, \bibinfo {author} {\bibfnamefont {R.~A.~M.}\
  \bibnamefont {Klomp}}, \bibinfo {author} {\bibfnamefont {M.~C.~M.}\
  \bibnamefont {Rentmeester}},\ and\ \bibinfo {author} {\bibfnamefont {J.~J.}\
  \bibnamefont {de~Swart}},\ }\bibfield  {title} {\bibinfo {title}
  {Partial-wave analysis of all nucleon-nucleon scattering data below 350
  mev},\ }\href {https://doi.org/10.1103/PhysRevC.48.792} {\bibfield  {journal}
  {\bibinfo  {journal} {Phys. Rev. C}\ }\textbf {\bibinfo {volume} {48}},\
  \bibinfo {pages} {792} (\bibinfo {year} {1993})}\BibitemShut {NoStop}%
%%CITATION = PHRVA,C48,792;%%
\bibitem [{\citenamefont {Bystricky}\ \emph {et~al.}(1978)\citenamefont
  {Bystricky}, \citenamefont {Lehar},\ and\ \citenamefont
  {Winternitz}}]{Bystricky:1976jr}%
  \BibitemOpen
  \bibfield  {author} {\bibinfo {author} {\bibfnamefont {J.}~\bibnamefont
  {Bystricky}}, \bibinfo {author} {\bibfnamefont {F.}~\bibnamefont {Lehar}},\
  and\ \bibinfo {author} {\bibfnamefont {P.}~\bibnamefont {Winternitz}},\
  }\bibfield  {title} {\bibinfo {title} {{Formalism of Nucleon-Nucleon Elastic
  Scattering Experiments}},\ }\href
  {https://doi.org/10.1051/jphys:019780039010100} {\bibfield  {journal}
  {\bibinfo  {journal} {J. Phys.(France)}\ }\textbf {\bibinfo {volume} {39}},\
  \bibinfo {pages} {1} (\bibinfo {year} {1978})}\BibitemShut {NoStop}%
%%CITATION = JOPQA,39,1;%%
\bibitem [{\citenamefont {La~France}\ and\ \citenamefont
  {Winternitz}(1980)}]{lafrance:jpa-00208966}%
  \BibitemOpen
  \bibfield  {author} {\bibinfo {author} {\bibfnamefont {P.}~\bibnamefont
  {La~France}}\ and\ \bibinfo {author} {\bibfnamefont {P.}~\bibnamefont
  {Winternitz}},\ }\bibfield  {title} {\bibinfo {title} {{Scattering formalism
  for nonidentical spinor particles}},\ }\href
  {https://doi.org/10.1051/jphys:0198000410120139100} {\bibfield  {journal}
  {\bibinfo  {journal} {{Journal de Physique}}\ }\textbf {\bibinfo {volume}
  {41}},\ \bibinfo {pages} {1391} (\bibinfo {year} {1980})}\BibitemShut
  {NoStop}%
\bibitem [{\citenamefont {Moravcsik}\ \emph {et~al.}(1989)\citenamefont
  {Moravcsik}, \citenamefont {Pauschenwein},\ and\ \citenamefont
  {Goldstein}}]{moravcsik:jpa-00210988}%
  \BibitemOpen
  \bibfield  {author} {\bibinfo {author} {\bibfnamefont {M.~J.}\ \bibnamefont
  {Moravcsik}}, \bibinfo {author} {\bibfnamefont {J.}~\bibnamefont
  {Pauschenwein}},\ and\ \bibinfo {author} {\bibfnamefont {G.~R.}\ \bibnamefont
  {Goldstein}},\ }\bibfield  {title} {\bibinfo {title} {{Amplitude systems for
  spin-1/2 particles}},\ }\href
  {https://doi.org/10.1051/jphys:0198900500100116700} {\bibfield  {journal}
  {\bibinfo  {journal} {{Journal de Physique}}\ }\textbf {\bibinfo {volume}
  {50}},\ \bibinfo {pages} {1167} (\bibinfo {year} {1989})}\BibitemShut
  {NoStop}%
\bibitem [{\citenamefont {Navarro~P{\'e}rez}\ \emph {et~al.}(2013)\citenamefont
  {Navarro~P{\'e}rez}, \citenamefont {Amaro},\ and\ \citenamefont
  {Ruiz~Arriola}}]{Perez:2013mwa}%
  \BibitemOpen
  \bibfield  {author} {\bibinfo {author} {\bibfnamefont {R.}~\bibnamefont
  {Navarro~P{\'e}rez}}, \bibinfo {author} {\bibfnamefont {J.~E.}\ \bibnamefont
  {Amaro}},\ and\ \bibinfo {author} {\bibfnamefont {E.}~\bibnamefont
  {Ruiz~Arriola}},\ }\bibfield  {title} {\bibinfo {title} {{Partial Wave
  Analysis of Nucleon-Nucleon Scattering below pion production threshold}},\
  }\href {https://doi.org/10.1103/PhysRevC.88.024002} {\bibfield  {journal}
  {\bibinfo  {journal} {Phys. Rev. C}\ }\textbf {\bibinfo {volume} {88}},\
  \bibinfo {pages} {024002} (\bibinfo {year} {2013})},\ \bibinfo {note}
  {[Erratum: Phys. Rev. C \textbf{88}, 069902 (2013)]},\ \Eprint
  {https://arxiv.org/abs/1304.0895} {arXiv:1304.0895} \BibitemShut {NoStop}%
%%CITATION = ARXIV:1304.0895;%%
\bibitem [{\citenamefont {{{Bayesian Analysis of Nuclear Dynamics (BAND)
  Framework project}}}(2020)}]{BAND_Framework}%
  \BibitemOpen
  \bibfield  {author} {\bibinfo {author} {\bibnamefont {{{Bayesian Analysis of
  Nuclear Dynamics (BAND) Framework project}}}}\ }(\bibinfo {year} {2020})\
  \bibinfo {note} {\url{https://bandframework.github.io/}}\BibitemShut
  {NoStop}%
\end{thebibliography}%

\end{document}